\documentclass[useAMS,usenatbib]{mnras}
\usepackage{amsmath}
\usepackage{graphicx}
\usepackage{txfonts}
\usepackage{bm}
\usepackage{longtable}
\usepackage{listings}

\def\apj{ApJ}
\def\apjl{ApJL}
\def\apjs{ApJ Suppl. Ser.}

\def\aap{A\&A}

\def\araa{ARAA}
\def\mnras{MNRAS}
\def\nat{Nature}

\def\beq#1{\begin{equation}\label{#1}}
\def\eeq{\end{equation}}
\def\beqa#1{\begin{eqnarray}\label{#1}}
\def\eeqa{\end{eqnarray}}

\def\Eq#1{Eq.~(\ref{#1})} 

\def\comment#1{\relax}

\newcommand{\msun}{M_\odot}

\title[Black hole spins in binary BH]{Black hole spins in coalescing binary black holes}
\author[Postnov \& Kuranov] {K.A. Postnov$^{1,2}$\thanks{E-mail: pk@sai.msu.ru},
A.G. Kuranov$^{1,3}$\\
$^1$ Sternberg Astronomical Institute, Moscow M.V. Lomonosov State University, Universitetskij pr., 13,  Moscow 119234, Russia\\
$^2$ Kazan Federal University, Kremlevskaya 18, 420008 Kazan, Russia\\
$^{3}$ Russian Foreign Trade Academy,  4a Pudovkin str., 119285 Moscow, Russia}

\begin{document}

\date{Received ... Accepted ...}
\pagerange{\pageref{firstpage}--\pageref{lastpage}} \pubyear{2018}

\maketitle

\label{firstpage}
\begin{abstract}

The possible formation mechanisms of massive close binary black holes (BHs) that can merge in the Hubble time to produce powerful gravitational wave bursts 
detected during advanced LIGO O1 and O2 science runs include the evolution from field low-metallicity massive binaries, 
the dynamical formation in dense stellar clusters and primordial BHs. Different formation channels produce different source distributions of total masses ${M_\mathrm{tot}}$ and effective spins $\chi_\mathrm{eff}$ of coalescing binary BHs.
Using a modified \textsc{bse} code, we carry out extensive population synthesis calculations of the expected effective spin and total mass distributions from the standard field massive binary formation channel for different metallicities of BH progenitors (from zero-metal Population III stars up to solar metal abundance), different initial rotations of the binary components, stellar wind mass loss prescription, different BH formation models and a range of common envelope efficiencies. The stellar rotation is treated in two-zone (core-envelope) approximation
using the effective core-envelope coupling time and with an account of the tidal synchronization of stellar envelope rotation during the binary system evolution. 
The results of our simulations, convolved with the metallicity-dependent star-formation history, show that the total masses and effective spins of the merging binary black holes detected during LIGO O1-O2 runs
but the heaviest one (GW170729) 
can be simultaneously reproduced by the adopted BH formation models. Noticeable effective spin of GW170729 requires additional fallback from the rotating stellar envelope. 

\end{abstract}


\section{Introduction}

The discovery of the first gravitational wave (GW) source GW150914 from coalescing binary black hole (BH) system  \citep{LIGO-PRL} not only heralded the beginning of gravitational wave astronomy era, but also stimulated a wealth of works on 
fundamental physical and astrophysical aspects of the formation and evolution of binary BHs. The LIGO binary BH detections   GW150914 \citep{LIGO-PRL}, GW151226 \citep{2016PhRvL.116x1103A}, 
GW170104 \citep{PhysRevLett.118.221101}, GW170608 \citep{2017arXiv171105578T} and recently announced additional binary BH coalescences \citep{LIGOO2},
as well as the first LIGO/VIRGO BH binary merging event GW170814 \citep{2017PhRvL.119n1101A}
enables BH masses and spins before the merging, 
the luminosity distance to the sources and the binary BH merging rate in the Universe to be estimated \citep{2016PhRvX...6d1015A}. Astrophysical implications of these measurements were discussed, e.g., in \cite{2016ApJ...833L...1A,2016ApJ...818L..22A}. 
This discovery was long awaited and anticipated from the
standard scenario of evolution of massive stars (see e.g. \citet{2018arXiv180605820M} for recent and \citet{2001PhyU...44R...1G} for early references).

The formation of double BHs from field stars is based on the evolution of single massive stars \citep{2002RvMP...74.1015W} and massive binary evolution scenario first proposed by \citet{1972NPhS..239...67V} and independently by \citet{1973NInfo..27...70T}. 
To produce a massive BH with $M\gtrsim 8-10M_\odot$ in the end of evolution, the progenitor star should have a large mass and low mass-loss rate at evolutionary stages preceding the core collapse. 
The mass-loss rate is strongly dependent on the metallicity, which plays the key role in determining the final mass of the 
stellar remnant (see e.g. \citealt{2010ApJ...714.1217B,2015MNRAS.451.4086S}).

In addition to the metallicity that affects the intrinsic evolution of the binary components, 
the most important uncertainty in the binary evolution is the efficiency of the common envelope (CE) stage which 
is required to form a compact double BH binary capable of merging within the Hubble time. 
In a dedicated study \citep{2016A&A...596A..58K}, high CE efficiencies ($\alpha_\mathrm{CE}< 1$) were found to be required 
for the possible formation of binary BH systems with parameters similar to GW150914 and GW151226 through the CE channel.
The common envelope efficiency remains a highly debatable issue. Recent model hydro simulations \citep{2016ApJ...816L...9O,2018arXiv181103656R}  failed to produce a high CE efficiency in both low-massive and massive binary stars, while successful CE calculations were reported by other groups (see, e.g., \cite{2016MNRAS.460.3992N}).  
It is not excluded that the so-called stable 'isotropic re-emission' mass transfer mode can be realized in high-mass X-ray binaries with massive BHs, thus helping to avoid 
the merging of the binary system components in the common envelope \citep{2017arXiv170102355V}. This stable mass transfer mode 
can explain the surprising stability of kinematic characteristics observed in the galactic microquasar SS433 \citep{2018MNRAS.479.4844C}.
Of course, much more
empirical constraints on and hydro simulations of the common evolution formation and properties are required.

To avoid the ill-understood common envelope stage, several alternative scenarios of the binary BH formation from massive stars were proposed. 
For example, in short-period  massive binary systems chemically homogeneous evolution due to rotational mixing can be realized. The stars remain
compact until the core collapse, and close binary BH system is formed without common envelope stage \citep{2016MNRAS.458.2634M,2016MNRAS.460.3545D,2016A&A...588A..50M}. 
In this scenario, a pair of nearly equal massive BHs can be formed with the merging rate comparable to the empirically inferred one from the first LIGO observations. This scenario, however, can be challenged by 
recent observations of slow rotation of WR stars in LMC \citep{2017arXiv170309857V}.

Another possible way to form massive binary BH system is through dynamical interactions in a dense stellar systems (e.g., globular clusters). This 
scenario was earlier considered by \cite{1993Natur.364..423S}. In the core of a dense globular clusters, stellar-mass BH form multiple systems, and 
BH binaries are dynamically ejected from the cluster. 
This mechanism can be efficient in producing 30+30 $M_\odot$ merging binary BHs \citep{2016ApJ...824L...8R,2017MNRAS.464L..36A}, and 
binary BH formed in this way can provide a substantial fraction of all binary BH mergings in the local Universe \citep{2016PhRvD..93h4029R,2018arXiv180901152R}. 

Finally, there can be more exotic channels of binary BH formation. For example, primordial black holes (PBHs) 
formed in the early Universe can form pairs which could be 
efficient sources of gravitational waves \citep{1997ApJ...487L.139N}. After the discovery of GW150914, the interest to binary PBHs
has renewed \citep{2016PhRvL.116t1301B}. Stellar-mass PBHs can form a substantial fraction of dark matter in the Universe \citep{2016PhRvD..94h3504C}. The PBHs formed
at the radiation-dominated stage can form pairs like GW150914 with the merging rate compatible with empirical LIGO results, being only some fraction of all 
dark matter \citep{2016arXiv160404932E,2016PhRvL.117f1101S}. A different class of PBHs with a universal log-normal mass spectrum produced in the frame of a modified Affleck-Dine supersymmetric baryogenesis mechanism \citep{ad-js,ad-mk-nk} were also shown to be able to match the observed properties of GW150914 \cite{2016JCAP...11..036B} without violating the existing constraints on stellar-mass PBH as dark matter (or at least its significant fraction). 

The aim of this paper is to calculate the total mass and effective spin distributions of coalescing binary black holes 
in the astrophysical scenario of BH-BH formation from initially massive binary stars. To do this, we use the population synthesis method based on the open \textsc{bse} (Binary Stellar Evolution) code elaborated in \cite{2000MNRAS.315..543H,2002MNRAS.329..897H}. The code was modified to take into account the BH formation from massive 
low-metallicity and zero-metallicity stars and was supplemented with a
treatment of the stellar core rotation during the evolution of massive stars in binaries.
In Section 2, we discuss the low effective spins inferred from GW observations of coalescing binary BHs. In Section 3, we describe modifications of the \textsc{BSE} code and the model assumptions used in the calculations. In Section 4, we describe in more detail the spin evolution of the binary components. Section 5 presents the results of our simulations, and Section 6 discusses and summarizes the main results. Examples of different types of calculated evolutionary tracks for several initial metallicites are given in the Appendix.

\section{Low effective spins of BHs in LIGO/VIRGO binary BH mergings}
\label{s:lowspins}

In General Relativity, a BH is fully characterized by its mass $M_\mathrm{BH}$ and dimensionless 
angular momentum $a=J/M_\mathrm{BH}^2$ (in geometrical units $G=c=1$) (the possible BH electric charge is negligible in real astrophysical conditions). 
The LIGO observations enable measurements of both masses of the coalescing BH components, $M_1$ and $M_2$, the total mass of the system, $M_\mathrm{tot}$, and the chirp mass that determines the strength of the gravitational wave signal ${\cal M}=(M_1M_2)^{3/5}/M_\mathrm{tot}^{1/5}$. From the analysis of waveforms at the inspiral stage, individual BH spins before the merging are poorly constrained, but their
mass-weighted total angular momentum parallel to the orbital angular momentum, $\chi_\mathrm{eff}$, can be estimated with acceptable accuracy 
\citep{2016PhRvX...6d1015A}. This parameter is $\chi_\mathrm{eff}=(M_1a_1\cos\theta_1+M_2a_2\cos\theta_2)/M_\mathrm{tot}$, where $\theta_i$ is the angle between the angular momentum of the i-th BH and orbital angular momentum of the binary system. The current LIGO/VIRGO detections suggest that observed merging events are consistent within measurements errors with $\chi_\mathrm{eff}\simeq 0$. The O2 LIGO event GW170104 may also have 
even slightly negative effective spin $\chi_\mathrm{eff}=-0.12^{+0.21}_{-0.30}$ \citep{PhysRevLett.118.221101}, suggesting that with a probability of around 0.8 the spin of one of the BHs prior to merging is directed by the angle more than 90 degrees relative to the orbital angular momentum of the binary system.

Here we should note that some uncertainties in the GW data analysis are still not excluded. For example, recently, the possibility of decreasing the BH masses by a factor of three compared to those as inferred from the GW signal analysis due to strong gravitational lensing was discussed by \cite{2018arXiv180205273B}. Also, an independent analysis of the reported
GW signal from GW150914 with an account of the waveform degeneracy from coalescing binary BHs \citep{2018arXiv180302350C} allows a possible increase in both masses and spins of the BH companions. Also, the LIGO analysis assumes random spin distributions of the components. Relaxing this prior could change the estimates (see, e.g., the discussion in \cite{2017arXiv170607053B}). Clearly, only future more precise observations can solve these open issues, and therefore below we will adopt the observational parameters of coalescing binary BHs as reported in the original LIGO/Virgo papers. 

Several explanations have been proposed to the low effective spins of the observed binary BH coalescences. For example, there can be a degeneracy between the eccentricity and spin corrections to the binary inspiral waveforms \citep{2017arXiv171106276H}. 
It is also possible that the BH spins can eventually lie in the binary orbital plane due to he dynamical evolution in triple systems, even for high initial BH spins coaligned with the orbital angular momentum \citep{2017arXiv171107142A}. 

The low effective BH spins as inferred from GW observations can have important evolutionary implications (see e.g. \cite{2016MNRAS.462..844K,2017arXiv170203952H,2017arXiv170607053B,2018PhRvD..97d3014W}). They suggest a slow rotation of the BH progenitors, which by itself strongly constrains, for example, chemically homogeneous pathways mentioned above in which the tidally induced rotation of the close binary  
components plays the key role. Massive stars are often observed to be rapid rotators \citep{2009pfer.book.....M}. No significant angular momentum loss is expected during
evolution of single stars with a low mass-loss rate by stellar wind and at the pre-collapse stage as required to produce massive BHs \citep{2015MNRAS.451.4086S}. 
Note that low effective spin values can imply either small intrinsic BH spins $a\sim 0$, or unusual orientations of BH spins with respect to the 
orbital angular momentum at the inspiral stage. The unusual spin orientations can, for example, be obtained in the dynamical formation scenario \citep{2016PhRvD..93h4029R}, where the BH spins are not expected to be correlated with the orbital angular momentum, or can result from natal BH kicks. The BH spin misalignment is also expected in merging BH binaries produced by Lidov-Kozai oscillations in triple stellar systems \citep{2018arXiv180503202L}. In the PBH scenario, BH spins must be intrinsically small as there are no vorticity 
in primordial cosmological perturbations. 

Therefore, the mass-spin distribution of BHs can serve as a sensitive tool to discriminate between different 
astrophysical formation channels of coalescing massive binary BHs \citep{2017arXiv170408370T,2018arXiv180503046N,2018arXiv180701336P}. 

\begin{figure}
\begin{center}
\includegraphics[width=0.49\textwidth]{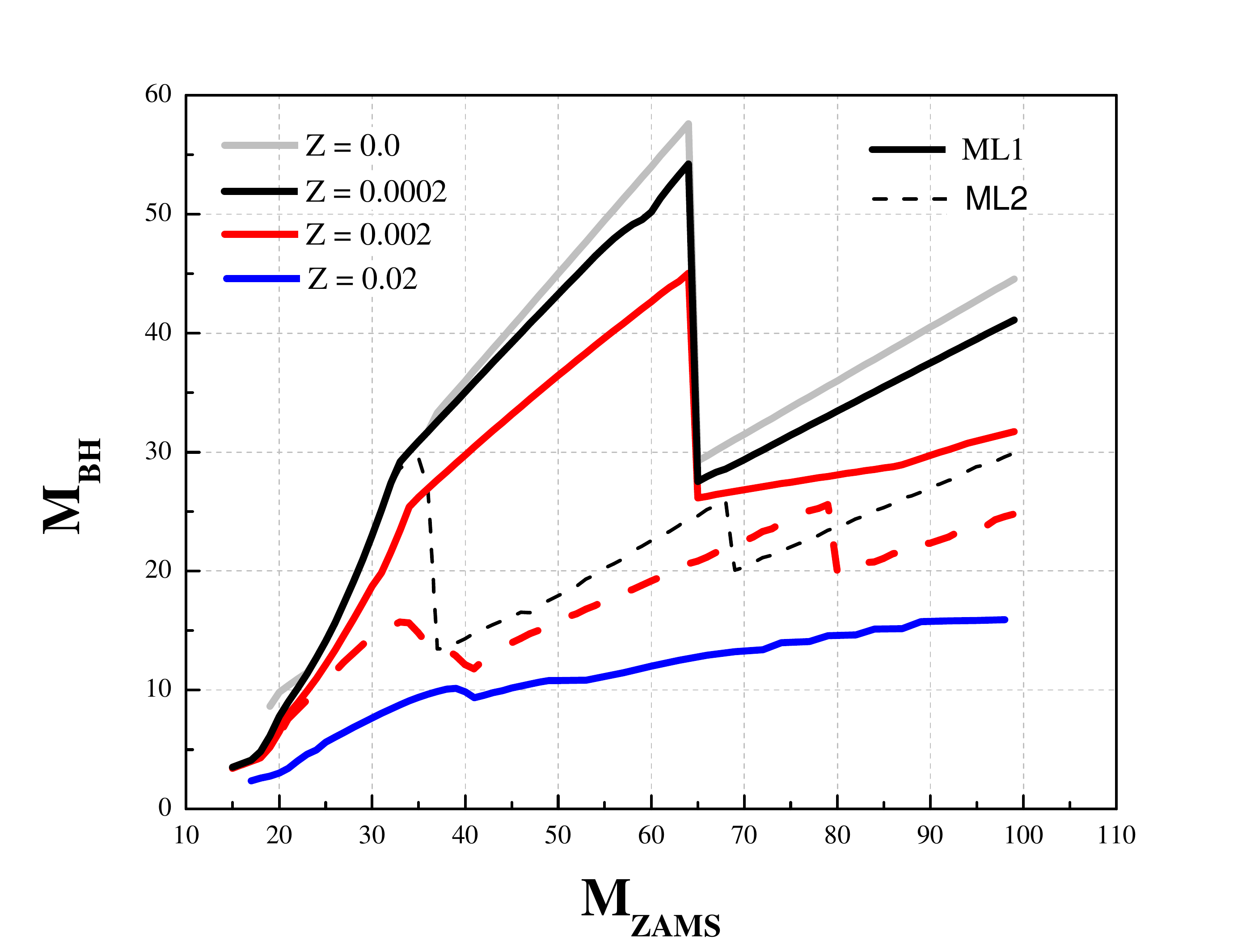}
\caption{BH remnant mass in the delayed core collapse mechanism \protect\citep{2012ApJ...749...91F} from stars of different metallicity for two stellar wind mass-loss models ML1 (\protect\cite{2018MNRAS.474.2959G}, the solid lines) and ML2 (\protect\cite{2001A&A...369..574V}, the dashed lines). For zero- and low-metallicity stars, the BH mass drop at around 60 $M_\odot$ is due to taking into account PPISN. For solar metal abundance (the bottom solid curve), the curves for Ml1 and ML2 wind mass-loss models are virtually indistinguishable.}
\label{f:mzams_mbh}
\end{center}
\end{figure}

\begin{figure*}
\begin{center}
\includegraphics[width=0.49\textwidth]{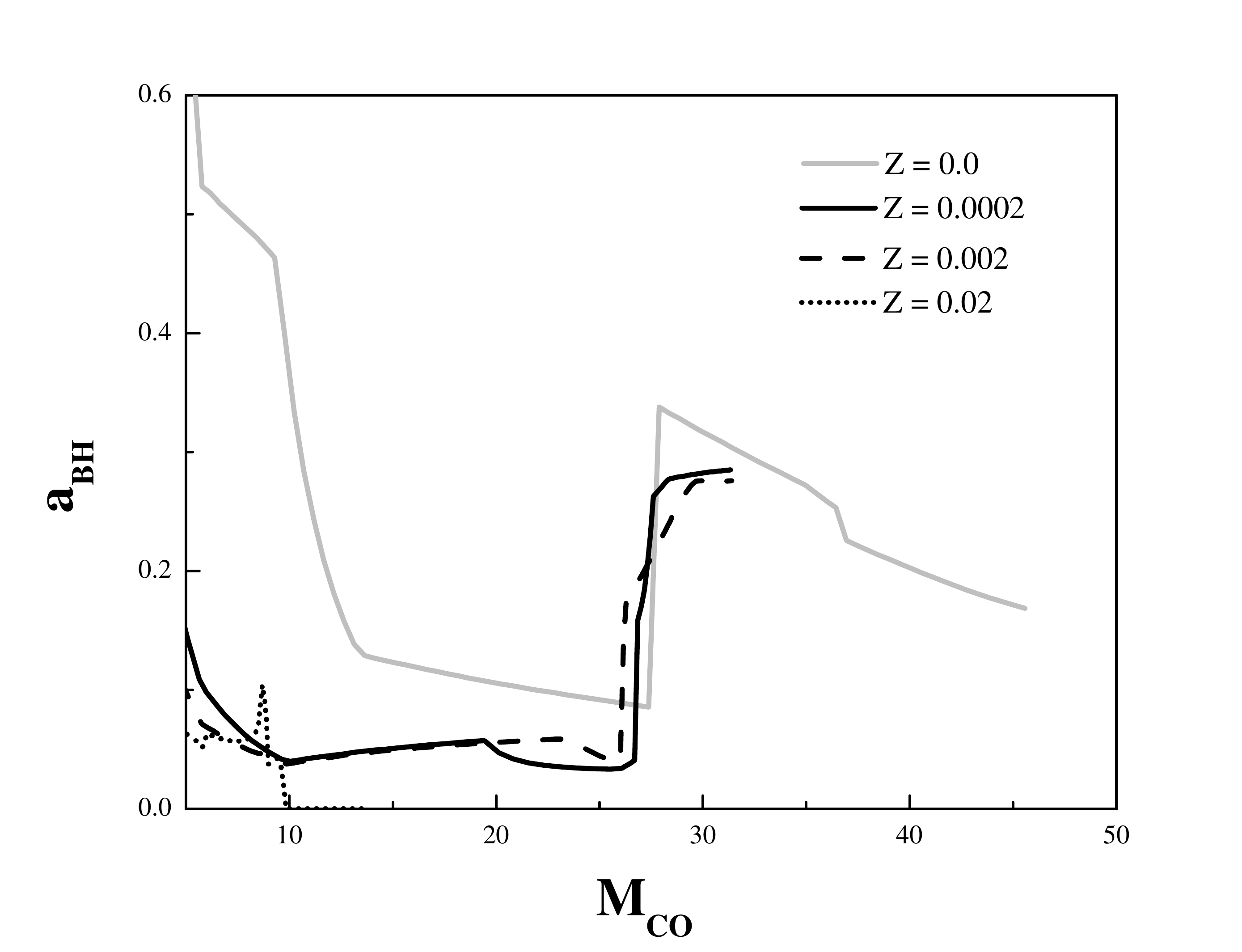}
\hfill
\includegraphics[width=0.49\textwidth]{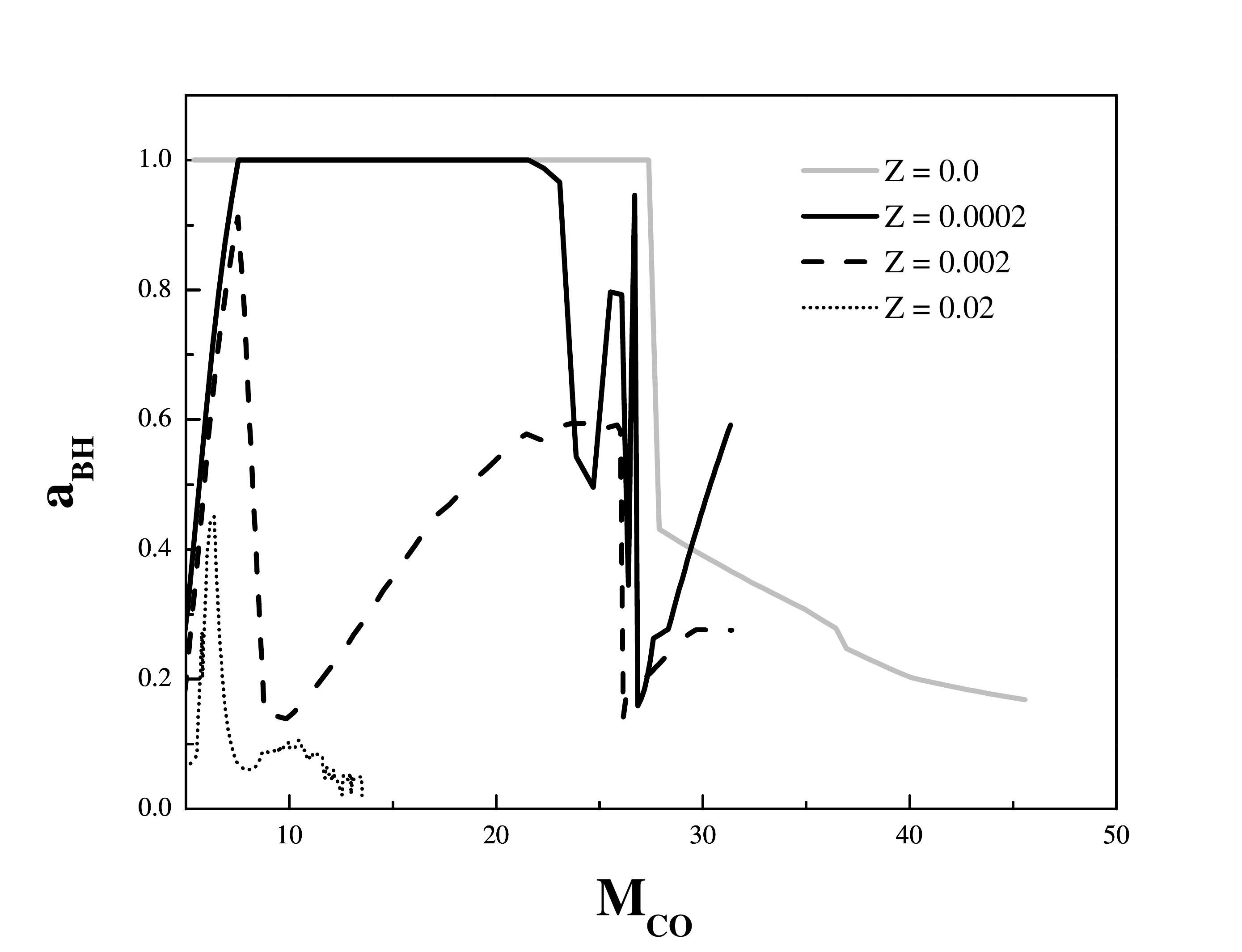}
\caption{Left: Spins of single black holes formed from the collapsing cores of massive rotating stars of different metallicity in the BH formation scenario with $M_\mathrm{BH}=0.9 M_\mathrm{CO}$, where  $M_\mathrm{CO}$ is the mass of the C-O core, as a function of the C-O core mass $M_\mathrm{CO}$. Right: Spins of single black holes assuming the  stellar envelope fallback ($M_\mathrm{BH}$ is calculated as in the delayed model by \protect\citealt{2012ApJ...749...91F}). The spin of the BH remnant is calculated using \Eq{e:a}.}
\label{f:core_spin}
\end{center}
\end{figure*}

\section{Modifications of the \textsc{bse} code}
\label{s:BSE}

The open-access \textsc{bse} code has been widely used by many authors to
make independent population synthesis calculations. The most recent modification was reported by \citep{2018MNRAS.474.2959G}. 
We added the code with the treatment of the evolution of zero-metallicity (primordial Population III) stars parametrized as in \cite{Kinugawa2014} 
and with the treatment of the rotation of the core of a star in a binary system using the effective core-envelope coupling time \protect\cite{2016MNRAS.463.1642P}, as described below.

\begin{itemize}
\item
The initial parameters of binaries with non-zero metal abundances are: the primary mass is distributed according to the Salpeter law, $dN/dM_1\propto M_1^{-2.35}$ ($0.1 M_\odot\le M_1\le 100 M_\odot$), the binary mass ratio $q=M_2/M_1\le 1$ is assumed to follow a flat distribution, $dN/dq= const$, the binary orbital separation are distributed following \cite{2012Sci...337..444S}, and a flat distribution for orbital eccentricities in the range [0,1]\footnote{The form of the initial binary eccentricity distribution is found to insignificantly affect final results.}. For zero-metallicity stars, we adopt different distributions as discussed in \citep{2017MNRAS.471.4702B} (model FS1 from Table 2 in that paper).

\item
Stellar wind mass loss is recognized to be one of the most important parameters that determines the mass and rotation of the stellar remnant. In the present calculations, we used the metallicity-dependent stellar wind mass loss from O-B stars with radiation pressure
corrections decreasing the stellar wind power as described in \citep{2018MNRAS.474.2959G} (model ML1 below), or without them (model ML2, \cite{2001A&A...369..574V}). Note that more massive BH remnants are produced in the ML1 stellar wind model than in the ML2 case (see Fig. \ref{f:mzams_mbh}). We assume no wind mass loss in zero-metal Population III stars.

\item
As the BH formation is not yet fully understood, to determine the mass of the BH remnant we have considered two cases: 

(i) The BH remnant with a mass equal to that of the pre-collapse C-O core of the BH progenitor (as calculated in the \textsc{bse} code), $M_\mathrm{BH}=0.9M_\mathrm{CO}$; the total mass of the coalescing
binary BH in this case is then (with an account of the 10\% gravitational mass defect) $M_\mathrm{tot}=0.9(M_\mathrm{CO,1}+M_\mathrm{CO,2})$.

(ii) The BH remnant with a mass $M_\mathrm{BH}$ as calculated by \cite{2012ApJ...749...91F} (the delayed model in that paper) and parametrized in the Appendix of \citep{2018MNRAS.474.2959G}. In this case,  
the mass of a BH is defined as:
\begin{equation}
M_{\mathrm{BH}} = 0.9(M_{\mathrm{Fe}} + \Delta M)
\end{equation}
where the mass of the proto-compact object (in fact, the iron stellar core) depends on the mass of the C-O core $M_{\mathrm{CO}}$:
\begin{equation}
M_{\mathrm{Fe}} = \begin{cases} 1.2\msun & \rm{if}~~ M_{\mathrm{core}}/\msun < 2.5  \cr
1.3\msun & \rm{if}~~ 3.5 \leq M_{\mathrm{CO}}/\msun < 6.0\cr
1.4\msun & \rm{if}~~ 6.0 \leq M_{\mathrm{CO}}/\msun < 11.0\cr
1.6\msun & \rm{if}~~ 11.0 \leq M_{\mathrm{CO}}/\msun. \cr
\end{cases}
\end{equation} 	
The additional matter falling on the collapsing iron core from the outer stellar envelope is:
\begin{equation}
\Delta M = \begin{cases} 0.2\msun & \rm{if}~~ M_{\mathrm{CO}}/\msun < 2.5  \cr
0.5M_{\mathrm{CO}}-1.05\msun & \rm{if}~~ 2.5 \leq M_{\mathrm{CO}}/\msun < 3.5\cr
(\alpha_{\rm D} M_{\mathrm{CO}} + \beta_{\rm D}) ({M_{\mathrm{fin}}- M_{\mathrm{Fe}}})& \rm{if}~~ 3.5 \leq M_{\mathrm{CO}}/\msun < 11.0\cr
 ({M_{\mathrm{fin}}- M_{\mathrm{Fe}}})& \rm{if}~~11.0 \leq M_{\mathrm{CO}}/\msun,  \cr 
\end{cases}
\end{equation} 
where $M_\mathrm{fin}$ is the total mass of the pre-collapse star (i.e., the core plus envelope), and 
\begin{equation}
\alpha_{\rm D} \equiv 0.133 - \frac{0.093}{M_{\mathrm{fin}}- M_{\mathrm{Fe}}}\,; \qquad
\beta_{\rm D} \equiv 1 - 11\alpha_{\rm D}~.
\end{equation}
The total mass of the coalescing binary BH is then $M_\mathrm{tot}=M_\mathrm{BH,1}+M_\mathrm{BH,2}.$
The pulsation pair instability expected in the helium cores of very massive stars $32 M_\odot\lesssim M_\mathrm{He}\lesssim 64 M_\odot$ \citep{2017ApJ...836..244W} is assumed to prevent the formation of BH remnants with masses above $\approx 52 M_\odot$.

\item
The BH remnant masses resulted from the delay core collapses SN mechanism for single stars 
with different metallicity are presented in Fig. \ref{f:mzams_mbh} as a function of the mass of the progenitor zero-age main-sequence (ZAMS) star for 
two adopted models of the stellar wind mass loss ML1 and ML2. The drop in the BH masses seen at around 60 $M_\odot$ is due to the pulsation pair-instability. The pair-instability supernovae (PISNe) and pulsation pair-instability supernovae (PPISNe) are treated using the formalism described in \citep{2018MNRAS.474.2959G}. 

\item
To calculate the spin of the BH remnant, we assume the angular momentum conservation of the collapsing C-O core. Therefore,
in the first BH formation scenario ($M_\mathrm{BH}=0.9M_\mathrm{CO}$), the BH spin is determined by the angular momentum of the collapsing C-O core only: $J_{BH}=J_{CO}$.
In the second case (the delayed core collapse mechanism),  the BH angular momentum 
increases due to the fallback of matter from the rotating outer envelope onto the C-O core:
$J_{BH}=J_{CO}+\Delta J_\mathrm{fb}$. Here 
\beq{e:jfb}
\Delta J_\mathrm{fb}=\Delta M_\mathrm{fb} j_\mathrm{fb}\,,
\eeq
where $\Delta M_\mathrm{fb}$ is the fallback mass, 
\beq{}
\Delta M_\mathrm{fb}=\max\{M_\mathrm{BH}-0.9M_\mathrm{CO},0\}
\eeq
which is the difference between the mass of the final BH  
and the mass of the collapsed C-O core. 
Below we shall refer to   
the BH formation models (i) and (ii) as the case without and with fallback from the envelope, respectively.

In the case (ii), we assume $j_\mathrm{fb}=\delta GM_\mathrm{BH}/c$ to be the mean specific angular momentum of
the matter falling onto BH from the rotating shell around the collapsing core. We set the dimensionless factor $\delta=2$, which is 
a compromise between the specific angular momentum of particles at the innermost stable orbit in the Schwarzschild metric ($\delta=2\sqrt{3}$) and 
in the extreme Kerr metric ($\delta=2/\sqrt{3}$). Such a choice is motivated by 
the fact that part of the envelope will accrete onto the BH through an accretion
disc, and some part can fall directly to the BH with a smaller specific angular momentum. Thus, the final dimensionless spin of the BH 
is defined as
\beq{e:a}
a=\min\left(1,\frac{(J_\mathrm{CO} + \min\{\Delta J_\mathrm{fb}, J_\mathrm{env}\})c}{GM_\mathrm{BH}^2}\right)\,
\eeq
where $G, c$ are the Newtonian gravitational constant and the speed of light, respectively, $J_\mathrm{CO}$ is the angular momentum of the C-O core which is assumed to be conserved during the gravitational collapse, $J_\mathrm{env}$ is the angular momentum of the stellar envelope prior to the core collapse.
The results of more detailed calculations can be found in the recent
paper by the Geneva group \citep{2018arXiv180205738Q}.

\item
In both BH formation scenarios (i) and (ii) we assume a natal BH kick ($V_{\rm kick}$):
\begin{equation}
\label{e:vkick}
V_{\rm kick} = \frac{M_\mathrm{fin}-M_\mathrm{BH}}{M_\mathrm{fin}-M_\mathrm{Fe}}\,{}A_{\rm kick},
\end{equation}
Here the amplitude $A_{\rm kick}$ is a Maxwellian distribution with a 1D-rms value $265$ km s$^{-1}$. 

\item
The common envelope phase is $\alpha$-parametrized: $\Delta E_\mathrm{env}=\alpha_\mathrm{CE}\Delta E_\mathrm{orb}$, where $\Delta E_\mathrm{orb}$ is the orbital energy of the binary lost in the common envelope stage, $\Delta E_\mathrm{env}$ is the binding energy of the envelope \citep{1984ApJ...277..355W,1984ApJS...54..335I}.   
To avoid the $\lambda$-description of the envelope binding energy, $\Delta E_\mathrm{env}=GM_\mathrm{env}M_\mathrm{core}/(\lambda R)$, we have directly 
calculated $\Delta E_\mathrm{env}$ using the open-access code described in \cite{2011ApJ...743...49L}.

\end{itemize}

\section{Spin evolution of the binary components}
\subsection{Core-envelope coupling}
\label{s:c-e}

As the effective 
spin $\chi_\mathrm{eff}$ of a coalescing BH-BH binary depends on the value and orientation of BH spins, 
we should specify how to calculate BH spins and their orientation relative to the binary orbital angular momentum. Here the following processes have been taken into account.

The value of a BH remnant spin $a$ depends on the rotational evolution of the stellar core, which is ill-understood and strongly model-dependent. 
For massive binaries, one possible approach is to match theoretical predictions of the core rotation with observed period distribution of young neutron stars observed as radio pulsars \citep{2016MNRAS.463.1642P}. Initially, a star is assumed to rotate rigidly, but after the main sequence the stellar structure can be separated in
two parts -- the core and the envelope, with some effective coupling between these two parts. 
In the present calculations, the separation of the star into the 'core' and the 'envelope' is done according to the scheme used in \cite{2000MNRAS.315..543H}.

The coupling between the core and envelope rotation can be mediated by magnetic dynamo \citep{2002A&A...381..923S}, internal gravity waves \citep{2015ApJ...810..101F}, etc. In this approximation, the time evolution of the angular momentum of the stellar core reads
\beq{e:c-e}
\frac{d\bf{J_c}}{dt}=-\frac{I_cI_e}{I_c+I_e}\frac{\bf{\Omega_c}-\bf{\Omega_e}}{\tau_c}\,,
\eeq
where $I_c$ and $I_e$ is the core and envelope moment of inertia, respectively,
calculated as in the \textsc{bse} code and $\bf{\Omega_c}$ and $\bf{\Omega_e}$ are their angular velocity vectors, which can be 
misaligned in due course of the evolution (see below). Long $\tau_c$ correspond to the case of an almost independent rotational evolution of the stellar core and the envelope, while short $\tau_c$ describes the opposite case of a very strong core-envelope rotational coupling. For the initially rigidly rotating single stars with $\Omega_\mathrm{c}=\Omega_\mathrm{e}$ this equation implies a slowing down of the core during the evolution because of the envelope radius increase and stellar wind mass loss. In the case of binary stars, the tidal effects can change $\Omega_\mathrm{e}$ differently, and the evolution of $\Omega_\mathrm{c}$ becomes more complicated (see below).

The validity of such an approach was checked by direct \textsc{MESA} calculations of the rotational evolution of a 15~$M_\odot$ star \citep{2016MNRAS.463.1642P}.
It was found that the observed period distribution of young pulsars can be reproduced if the effective coupling time between the core and envelope of a massive star is $\tau_c=5\times 10^5$~years (see Fig. 1 in \cite{2016MNRAS.463.1642P}). Below we shall assume that this parametrization of the core-envelope angular momentum coupling is also applicable to the evolution of very massive stars leaving behind BH remnants. In our calculations,  we varied the value of $\tau_c$ from $10^4$ years (very strong coupling) to $10^7$ years (very weak coupling).

\subsection{Initial rotational velocity of the components}

The initial rotational velocity of the binary components was chosen according
to the empirical relation between the mass of main-sequence stars $M_0$  and their
equatorial velocities (as used, e.g., in the \textsc{bse} code \citep{2002MNRAS.329..897H})
\beq{e:vrot}
v_{0}=330\frac{M_0^{3.3}}{15+M_0^{3.45}}\hbox{km\,s}^{-1} 
\eeq
(here $M_0$ is in solar units). 
The main-sequence stars were assumed to be initially uniformly rotating. This assumption 
has some support from \textit{Kepler} asteroseismology
\citep{2016arXiv161203092M}.

To check the effect of the initial rotational velocity, we performed 
calculations for (a) initially non-rotating stars, $v_\mathrm{rot}=0$, (b) stars rotating according to \Eq{e:vrot}, 
$v_\mathrm{rot}=v_0$, and
(c) stars rotating with $v_\mathrm{rot}=\min(4v_0, v_\mathrm{crit})$, where $v_\mathrm{crit}=\sqrt{(2/3)GM_0/R_0}$ is the 
limiting equatorial (break-up) velocity for a rigidly rotating star with mass $M_0$ and
polar radius $R_0$.

\subsection{Components spin alignment/misalignment}

The initial spins of the components of a binary system are likely to be coaligned with the orbital angular momentum $\bf{\hat L}$. This assumption is
supported by recent observations of coaligned spins of stars in old stellar clusters 
\citep{2017NatAs...1E..64C}. However, due to violent turbulence in proto-stellar clouds and possible dynamical interactions, spins of the binary component (especially for 
sufficiently large orbital separations) can be initially misaligned. 
The latter possibility is supported by observations of misaligned protostellar and protoplanetary discs in binary systems
(see, e.g., observations of HK Tau \cite{2014Natur.511..567J}, IRS 43 \citep{2016ApJ...830L..16B}), which can be explained by the binary formation in the turbulent fragmentation process (e.g. \cite{2016ApJ...827L..11O}).
Therefore, in our calculations we will consider two extreme cases: (i) the initial spins of the binary components aligned with the orbital 
angular momentum and (ii)
totally independent (random) initial spin orientation of the binary components. 

In the course of the binary evolution, the spin-orbit misalignment
can be also produced by an additional kick velocity during the BH formation
\citep{1999astro.ph..3193P,2000ApJ...541..319K,2001PhyU...44R...1G}. The possibility of BH generic kicks is actively debated in the literature; see, e.g., 
the recent discussion of potential constraining BH natal kicks from GW observations in \citep{2017arXiv170403879O,2017arXiv170407379Z,2018PhRvD..97d3014W}. 
In our calculations, we adopted the fallback-dependent BH kicks described by \Eq{e:vkick}.

\subsection{Tidal synchronization of the envelope rotation and orbital circularization}

During the evolution of a binary system, we assume that the rotation of the stellar envelope  gets tidally synchronized with the orbital motion with the characteristic synchronization time $t_\mathrm{sync}$, and the processes of tidal synchronization and orbital circularization are treated as in the \textsc{bse} code (see \cite{2002MNRAS.329..897H}, Eqs. (11), (25), (26), (35)). Due to a possible misalignment of the spin vectors of the stars with the binary orbital angular momentum $\bf{\hat L}$ as discussed above, we separately treated the change of the core and the envelope spin components parallel and perpendicular to $\bf{\hat{L}}$, $\bf{J_{c,e}}=\bf{J_{||(c,e)}} +\bf{J_{\perp(c,e)}}$. On evolutionary stages prior to the compact remnant formation, for each binary component we assumed that due to the tidal interactions the stellar envelope spin components $J_{e(||, \perp)}$ evolve with  the characteristic time $t_\mathrm{sync}$:
\beq{}
\left.\frac{d\bf{J_e}}{dt}\right|_{tid}=I_e\left.\frac{d\bf{\Omega_e}}{dt}\right|_{tid}\,,
\eeq
where  the parallel and perpendicular to the orbital angular momentum components of the envelope angular velocity change as
\beq{e:o||}
\dot \Omega_{e,||}^{tid}=-\frac{\Omega_{e,||}-\Omega_\mathrm{orb}}{t_\mathrm{sync}}
\eeq
and
\beq{e:oper}
\dot \Omega_{e,\perp}^{tid}=-\frac{\Omega_{e,\perp}}{t_\mathrm{sync}}\,,
\eeq
respectively. Here we have assumed that the tidal interactions tend to exponentially synchronize the envelope's parallel rotation with the orbital motion (\Eq{e:o||}), and to decrease the perpendicular component of the envelope's rotation (\Eq{e:oper}) on the same time scale $t_\mathrm{sync}$. Clearly, under this assumption the spin-orbit alignment time is different from $t_\mathrm{sync}$. This time is also model-dependent. In our calculations, we used Eqs. (25) and (26)  from \cite{2002MNRAS.329..897H} 
for the tidal circularization and synchronization time, respectively. As the tidal interaction can be not as effective as adopted in that paper (see, e.g., \cite{2007A&A...467.1389}), we repeated calculations with the circularization and synchronization times multiplied by factor 100. No significant difference in the final results were found because of very efficient tidal effects at the stage of the Roche lobe overflow even with the increased characteristic times.

\begin{figure*}
\includegraphics[width=0.8\textwidth]{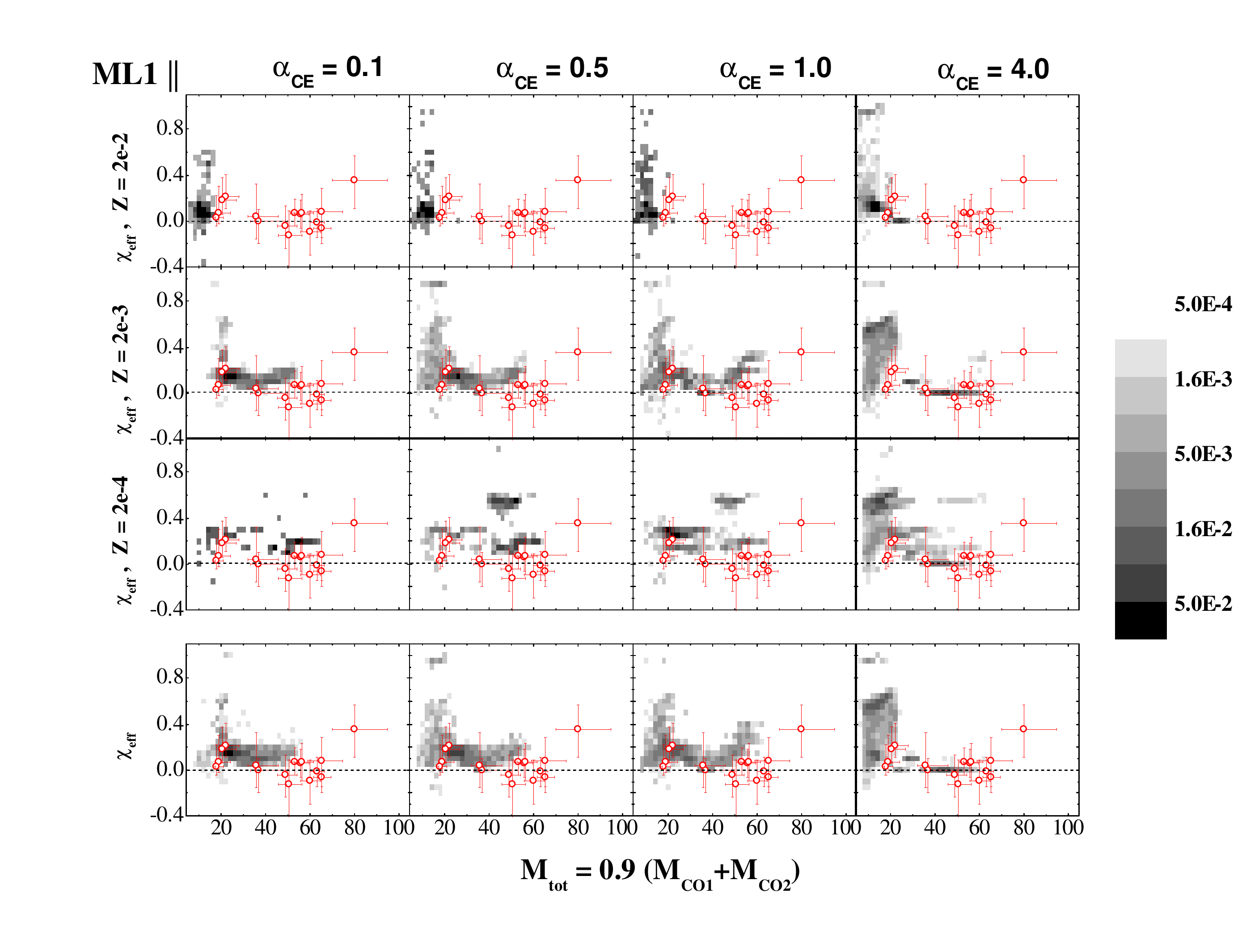}
\hfill
\includegraphics[width=0.8\textwidth]{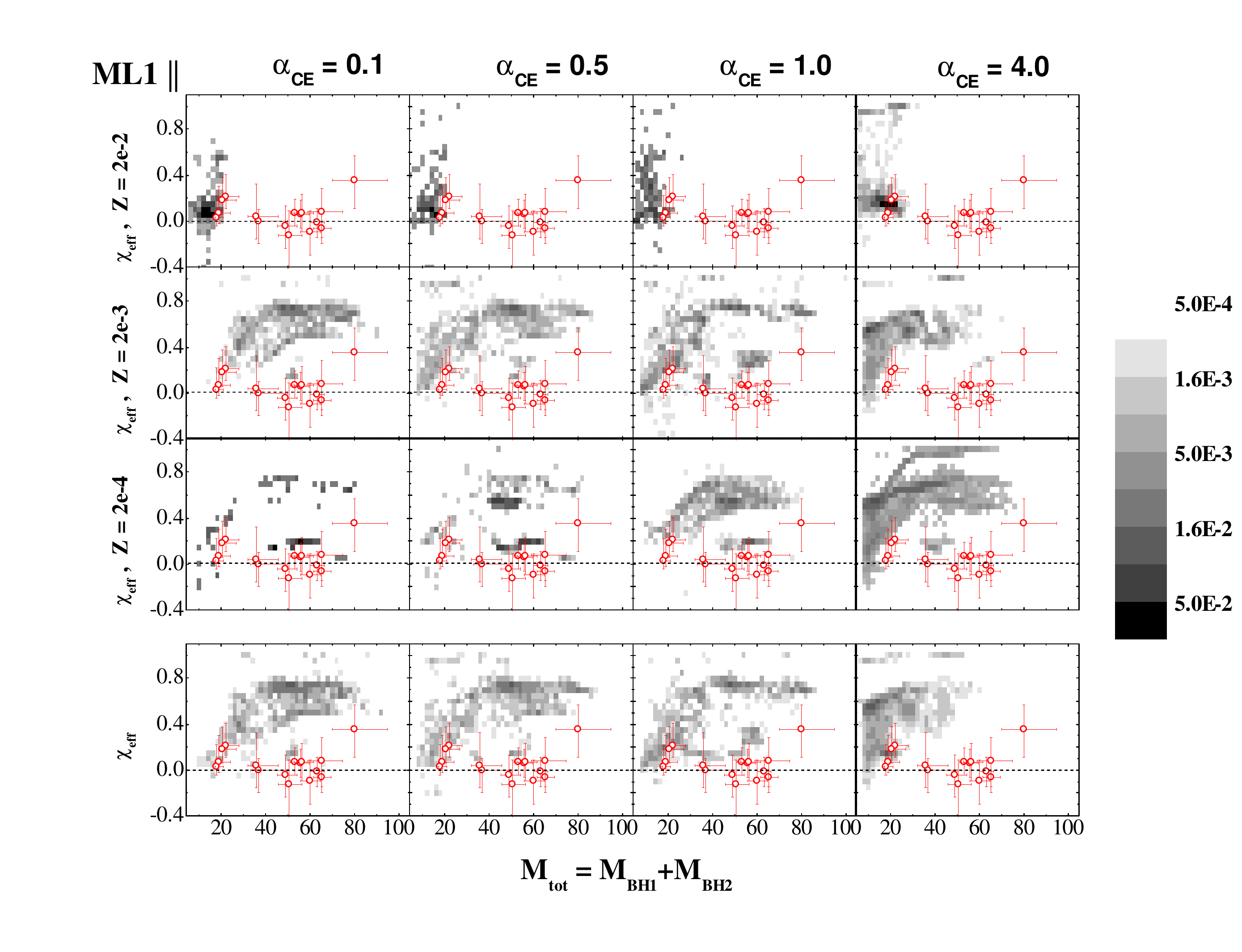}
 \caption{The normalized total mass -- effective spin ${M_\mathrm{tot}}-\chi_\mathrm{eff}$ distribution for the coalescing black hole binaries for different stellar metallicities (1st-3d row), the assumed metallicity-dependent star-formation rate history (\protect\cite{2018arXiv180707659E}, 4th row),  and different CE efficiencies (1st-4th column). The less effective stellar wind mass loss with radiation pressure corrections (model ML1, \protect\cite{2018MNRAS.474.2959G}) is assumed. The effective core-envelope coupling time is $\tau_c=5\times 10^5$ years.
 The initial binary component spins are coaligned with the orbital angular momentum.  The natal BH kick is given by \Eq{e:vkick}. 
Open circles with error bars show the observed BH-BH systems from LIGO GWTC-1 catalogue
\protect\citep{LIGOO2}.
Upper panel: BH formation model without the fallback from the stellar envelope, $M_\mathrm{tot}=0.9(M_\mathrm{CO,1}+M_\mathrm{CO,2})$.  Lower panel: 
BH formation model with the fallback from the envelope, $M_\mathrm{tot}=M_\mathrm{BH,1}+M_\mathrm{BH,2}$. }
   \label{f:JL0_MLW1}
\end{figure*}
\begin{figure*}
\begin{center}
\includegraphics[width=0.8\textwidth]{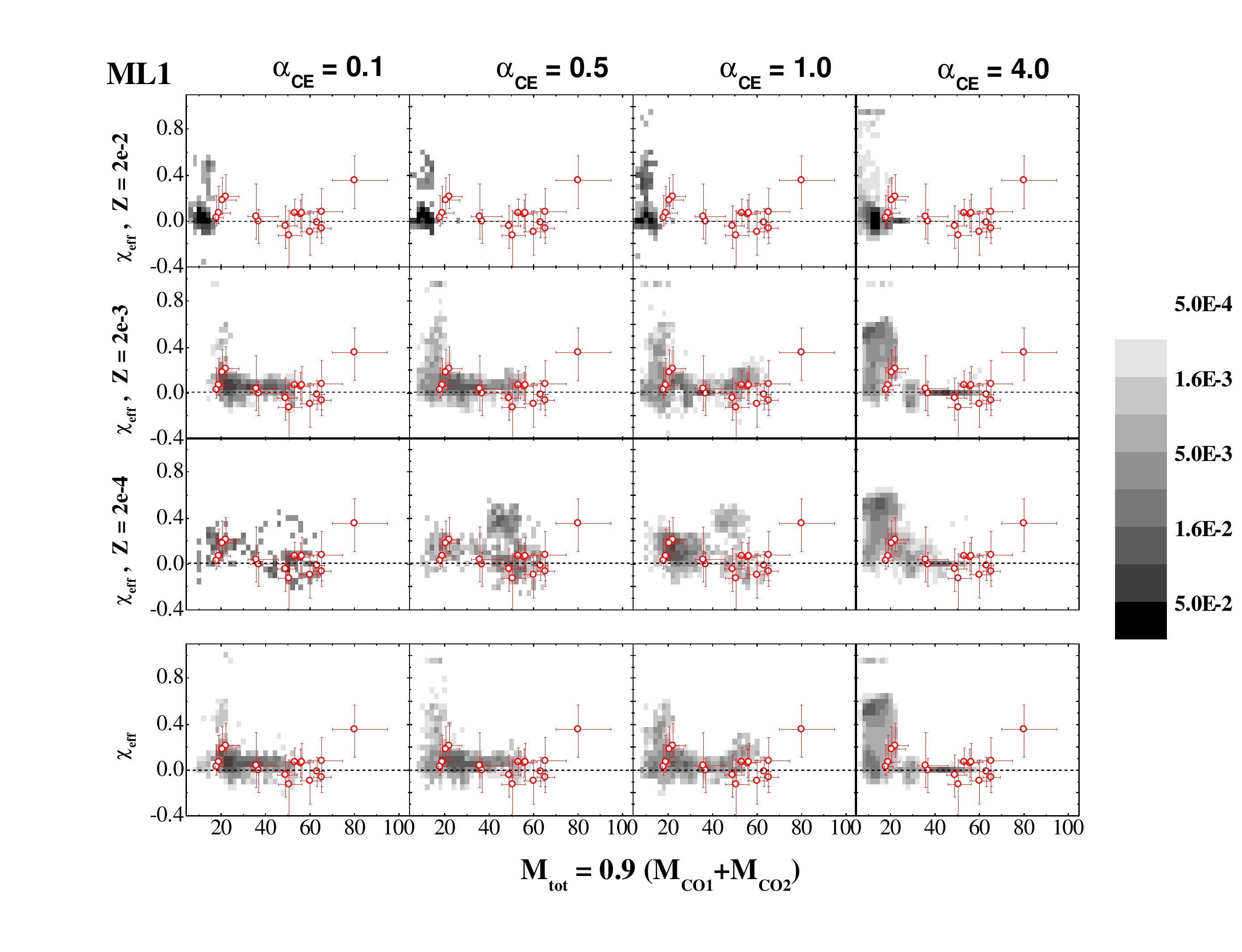}
\hfill
 \includegraphics[width=0.8\textwidth]{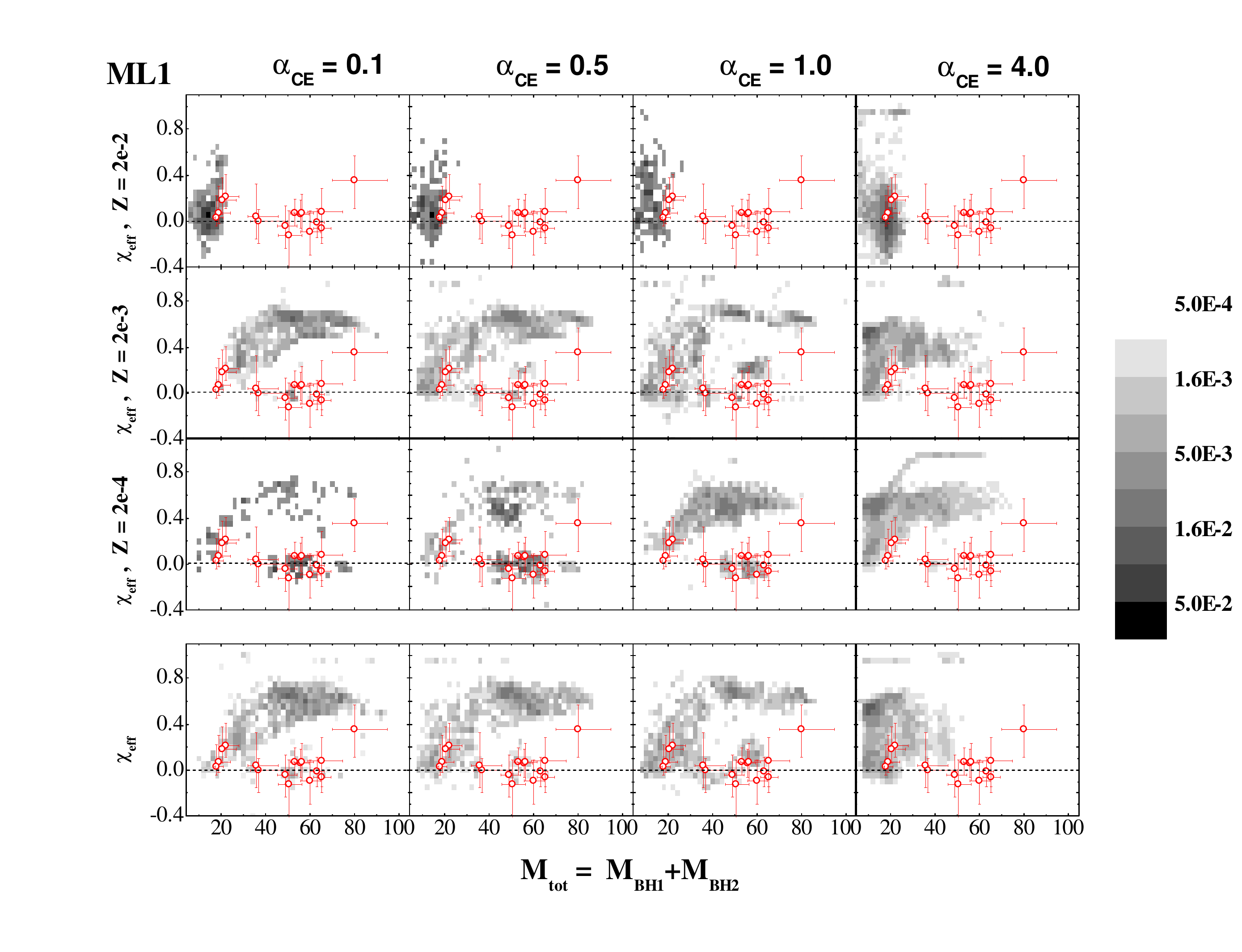}
\caption{The same as in Fig. \ref{f:JL0_MLW1} for the initial binary component spins randomly oriented relative to the orbital angular momentum. 
}
   \label{f:random_MLW1}
\end{center}
\end{figure*}
\begin{figure*}
\includegraphics[width=0.8\textwidth]{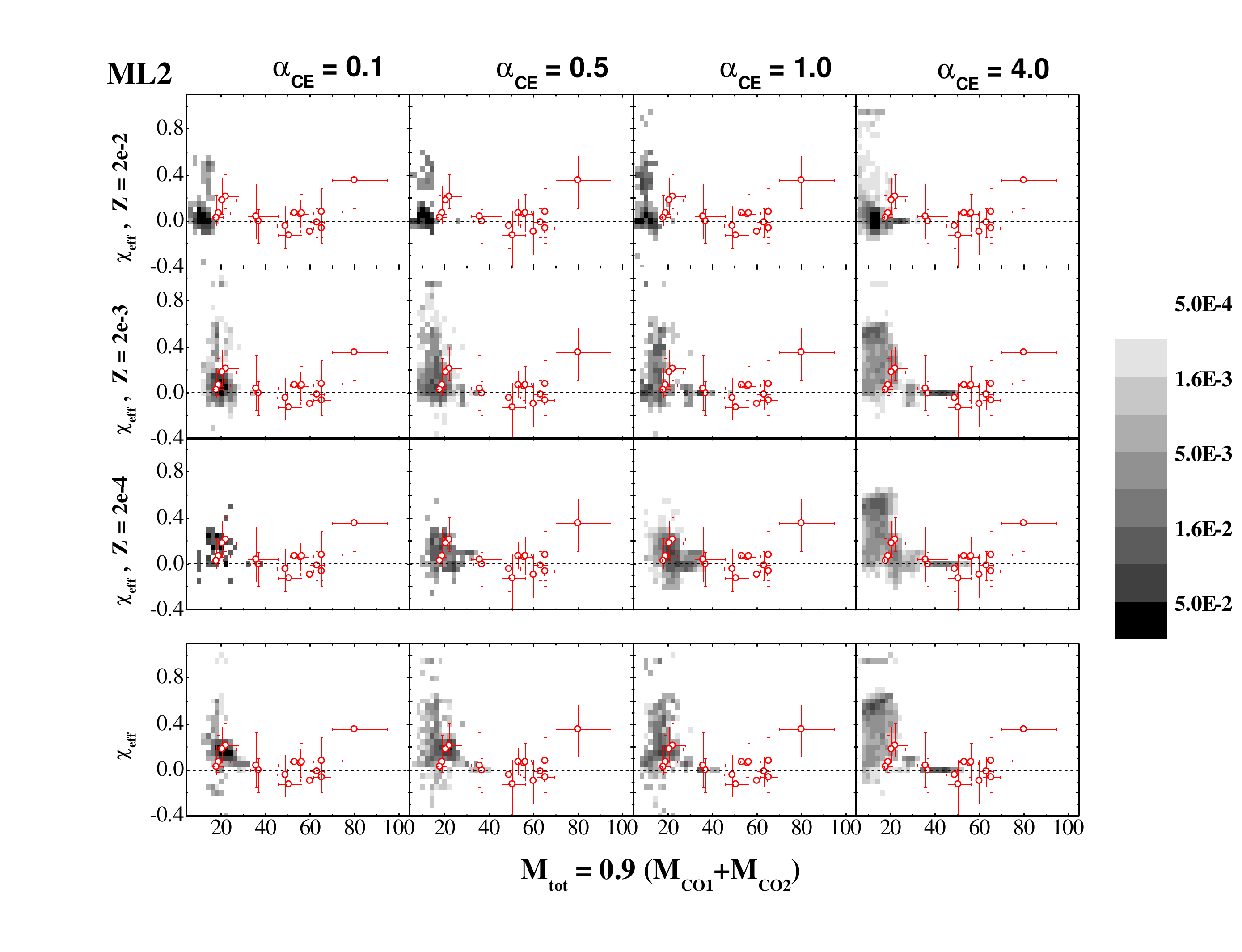}
\hfill
\includegraphics[width=0.8\textwidth]{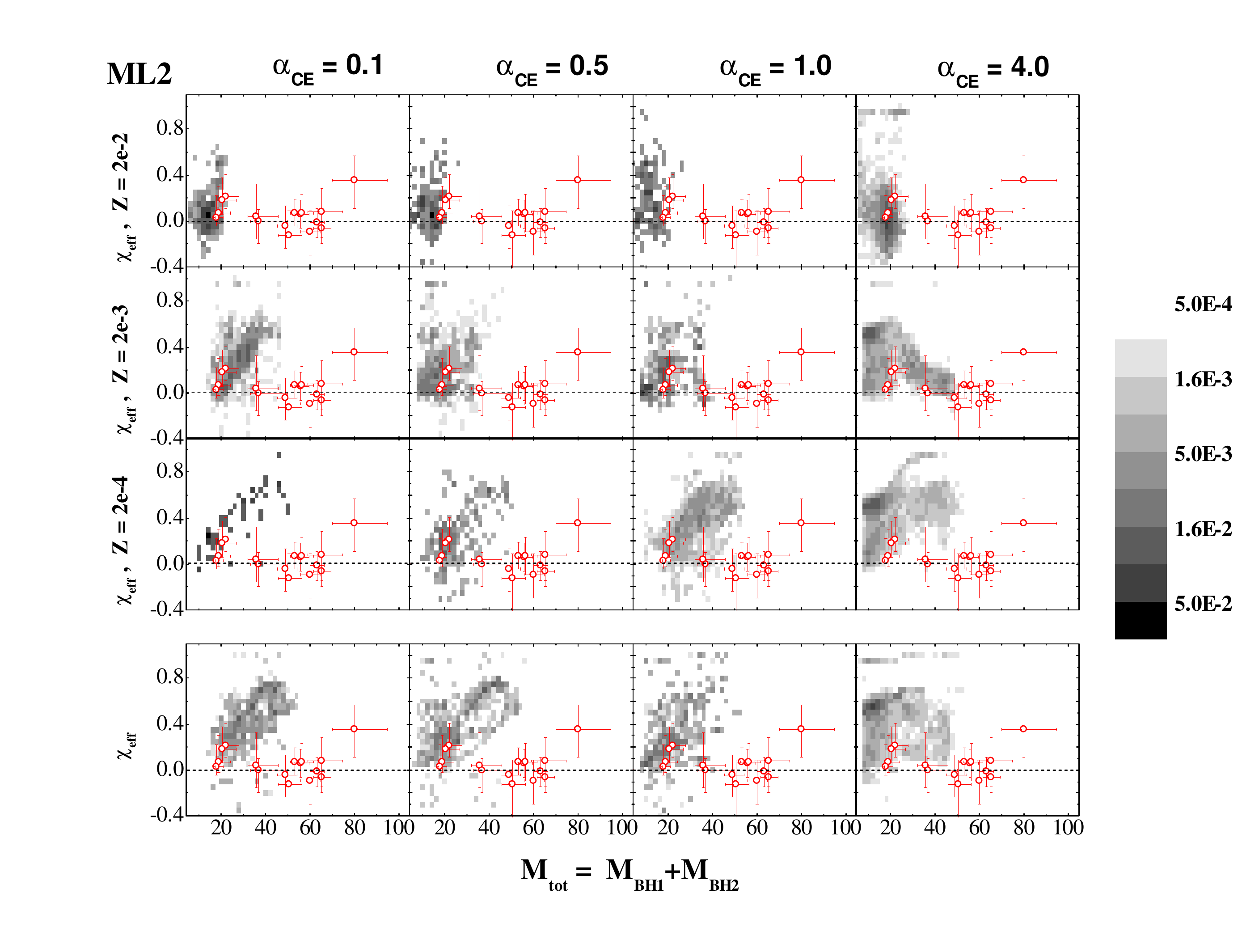}
 \caption{The same as in Fig. \ref{f:random_MLW1} for the more 
 effective stellar wind mass loss model ML2 \citep{2001A&A...369..574V}.
 }
   \label{f:random_MLW2}
\end{figure*}

\begin{figure*}
\includegraphics[width=0.8\textwidth]{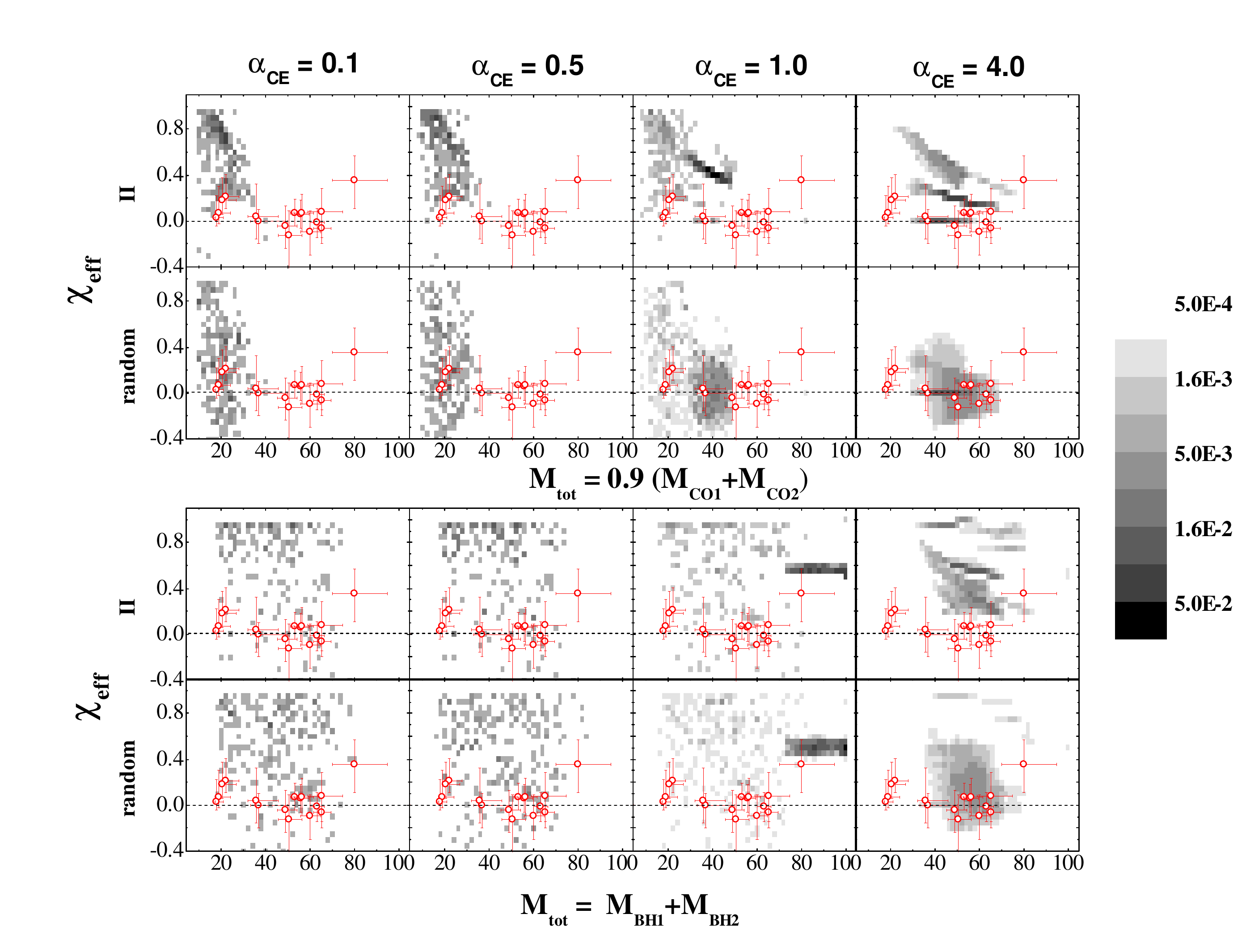}
 \caption{The same as in Fig. \ref{f:JL0_MLW1} for zero-metal Pop III stars \citep{Kinugawa2014} assuming no mass-loss. Upper and bottom panels show the results for the
 BH formation without and with fallback from the envelope, respectively, for the initially coaligned (upper rows) and randomly misaligned (bottom rows) binary component spins.
 }
   \label{f:Z0}
\end{figure*}

The spin of the stellar 
envelope, $\bf{J_{e}}$, also changes due to the core-envelope interaction with the characteristic time $\tau_c$ (see above, Section \ref{s:c-e}), the mass loss (mass gain) due to the stellar wind losses and the mass exchange between the components:
\beq{}
\frac{d\bf{J_e}}{dt}=-\frac{d\bf{J_c}}{dt}+\left.\frac{d\bf{J_e}}{dt}\right|_{tid}+\left.\frac{d\bf{J_e}}{dt}\right|_{\dot M}
\eeq
The spin evolution of the binary components described above 
was added to the updated \textsc{bse} population synthesis code.

\section{Results of simulations}

With the additions to the \textsc{bse} code as described above, a population synthesis of typically 1,000,000 binaries per run has been carried out for different parameters of binary evolution (the common envelope stage efficiency $\alpha_\mathrm{CE}$, stellar metallicities, stellar wind model) and assuming coaligned or misaligned initial spins of the binary components.

To compare the results of simulations with the BH masses and effective spins as inferred from the gravitational-wave observations, we need to take into account  the time delays from the formation and coalescence of a given binary system and the history of the star-formation rate in the Universe as a function of time. We used the method described in detail, e.g., in 
\cite{2015ApJ...806..263D,2016ApJ...819..108B}, with the star-formation dependence on the metallicity from
\cite{2018arXiv180707659E}. 
The fractional mass density of star formation at and below metallicity mass fraction of $Z$ at given redshift $z$ is factorized as
\beq{e:sfr1}
\Psi\left(z,\frac{Z}{Z_{\odot}}\right)=\psi(z)\Phi(Z/Z_\odot)
\eeq
where the cumulative metallicity distribution is
\beq{e:PhiZ}
\Phi(Z/Z_\odot)=
\frac{\hat{\Gamma}[0.84,(Z/Z_{\odot})^2 \, 10^{0.3 z}]}{\Gamma(0.84)},
\eeq
and $\hat{\Gamma}$ and $\Gamma$ are the incomplete and complete Gamma functions, respectively. The  $\psi(z)$ is star-formation rate density as a function of redshift $z$:
\beq{e:sfr2}
\psi(z) = 0.015 \frac{(1+z)^{2.7}}{1+((1+z)/2.9)^{5.6}} \, M_{\odot}\, {\rm yr^{-1} \, Mpc^{-3}}.
\eeq
The time-redshift relation is calculated using the standard $\Lambda$CDM cosmological model with $H_0=70$~km~s$^{-1}$~Mpc$^{-1}$, $\Omega_\Lambda=0.7$ and $\Omega_m=0.3$.
We have carried out simulations of stellar populations with the metallicity ranged from $Z_{mi}=10^{-4}$ to $Z_{max}=2\times 10^{-2}$ binned in 10 intervals in each decade so that the probability to find a system with given $Z_i$ at redshift $z$ is $P(Z_i)=(\Phi(Z_i+\Delta Z)-\Phi(Z_i))/(\Phi(Z_{max})-\Phi(Z_{min})[\Delta Z/(Z_{max}-Z_{min})]$. The calculated delay-time distribution of coalescing binary BHs produced by systems with given metallicity is convolved with the adopted star-formation rate redshift and metallicity distribution. We have used 100 ml yrs time intervals for the redshift convolution.

The horizon distance for aLIGO detector for a binary system with the chirp mass ${\cal{M}_c}$ is taken to be $D_h\simeq 450\mathrm{Mpc}\,({\cal{M}}_c/1.2 M_\odot)^{5/6}$. Corrections to the non-spherical response function of the detector are neglected. While the detector response function affects the detector horizon for a given ${\cal{M}}_c$, its effect is found to be subdominant compared to the convolution with the star-formation rate history. Besides, for 
coalescing BH binaries with noticeable (and possibly misaligned) spins, there are uncertainties in waveforms caused by the BH spin values and misalignment which require dedicated studies (cf. the waveform effects for non-spinning binaries shown in 
Figs.3-4 in \cite{2015ApJ...806..263D}).
Our main results are presented in Figs. \ref{f:JL0_MLW1}-\ref{e:vkick}.

\subsection{Effect of the fallback from the envelope during BH formation }

The possible fallback from the rotating stellar envelope onto a BH formed 
from the collapsing stellar core is found to mostly affect the 
distribution of coalescing binary black holes on the total mass -- effective spin $M_\mathrm{tot}-\chi_\mathrm{eff}$ plane (see Figs.
\ref{f:JL0_MLW1}-\ref{f:random_MLW2}). 
In these Figures, the observed BH-BH binaries from LIGO/Virgo GWTC-1 catalogue \citep{LIGOO2}
are shown by open circles with error bars in the order of decreasing total mass $M_\mathrm{tot}$ for the guidance.

The upper panel of 
Fig. \ref{f:JL0_MLW1} shows the results of calculations for different
common envelope efficiencies $\alpha_\mathrm{CE}$ (1st-4th columns for $\alpha_\mathrm{CE}=0.1, 0.5, 1, 4$, respectively ), the initial metallicity of the binary components (1st-3d row for $Z=0.02, 0.002, 0.0002$, respectively), and for the convolution of the results for different metallicites with the adopted star-formation rate history in the Universe \citep{2018arXiv180707659E} (4th row).
The results are shown for the stellar wind mass loss with radiation pressure corrections (model ML1, \cite{2018MNRAS.474.2959G}) and the BH formation model without fallback from the outer envelope,    
$M_\mathrm{tot}=0.9(M_\mathrm{CO,1}+M_\mathrm{CO,2})$. The initial spins of the binary components are coaligned with the orbital angular momentum, so that the negative effective spins of the coalescing BH systems arise solely due to the natal BH kicks (see \Eq{e:vkick}). 

The bottom panel
of Fig. \ref{f:JL0_MLW1} shows the results  obtained under the same assumptions as those presented in the upper panel but for the
BH formation model with fallback from the outer envelope, 
$M_\mathrm{tot}=M_\mathrm{BH,1}+M_\mathrm{BH,2}$. Particular types of evolutionary tracks leading to merging binary BH are summarized in the Appendix. 

The comparison of the 4th rows in the
upper and lower panels of Fig. \ref{f:JL0_MLW1} suggests that within measurement errors, the  
observed BH-BH systems but the most massive (GW150914,GW170729)  can be reproduced in the standard scenario for the adopted metallicity-dependent star-formation rate history \citep{2018arXiv180707659E} if no significant fallback from the rotating envelope is assumed. Adding the envelope fallback (the lower panel) leads to the appearance of rapidly rotating BH components (cf. Fig. \ref{f:core_spin}, right panel) and, consequently, of the coalescing BH binaries with high effective spins $\chi_\mathrm{eff}$ (e.g. GW170729). 
The systems with higher $M_\mathrm{tot}$ are also obtained in the case with the envelope fallback.

Fig. \ref{f:random_MLW1} shows the results of calculations assuming random initial orientations of the binary component spins. Generally, the results are similar to those 
shown in Fig. \ref{f:JL0_MLW1}, with a smoother distribution of $\chi_\mathrm{eff}$ and 
higher probability to have a negative effective spin prior to the coalescence, $\chi_\mathrm{eff}<0$, for some combination of the parameters. 
Like in the case of the coaligned initial spins, the addition of the envelope mass and angular momentum to the nascent BH (the bottom panel of Fig. \ref{f:random_MLW1}) gives rise to coalescing binary BHs with high effective spins (like GW170729).

For comparison, we have run calculations by assuming the more effective stellar wind mass-loss from massive stars (model ML2, \cite{2001A&A...369..574V}). The results remain qualitatively the same as in calculations with less effective stellar wind mass-loss model ML1 shown in Fig. \ref{f:random_MLW1} and are presented in Fig. \ref{f:random_MLW2} for randomly directed initial binary spins. As expected, somewhat lighter BH remnants are produced for a more effective wind mass-loss rate.

For completeness, in Fig. \ref{f:Z0} we present the result of calculations of
the evolution of the zero-metallicity Pop III stars using the model by \cite{Kinugawa2014} after a delta-like SFR burst. The total masses and effective spins of binary BHs produced from these stars can spread over a wide range covering the observed binary BH parameters. However, the evolution of 
these objects is  less certain and model-dependent, and
the coalescing binary BHs from Pop III stars are recognized to be subdominant to produce the observed binary BH coalescences (e.g., \cite{2016MNRAS.460L..74H,2016MNRAS.461.3877D,2017MNRAS.471.4702B}).

\begin{figure*}
\begin{center}
\includegraphics[width=0.49\textwidth]{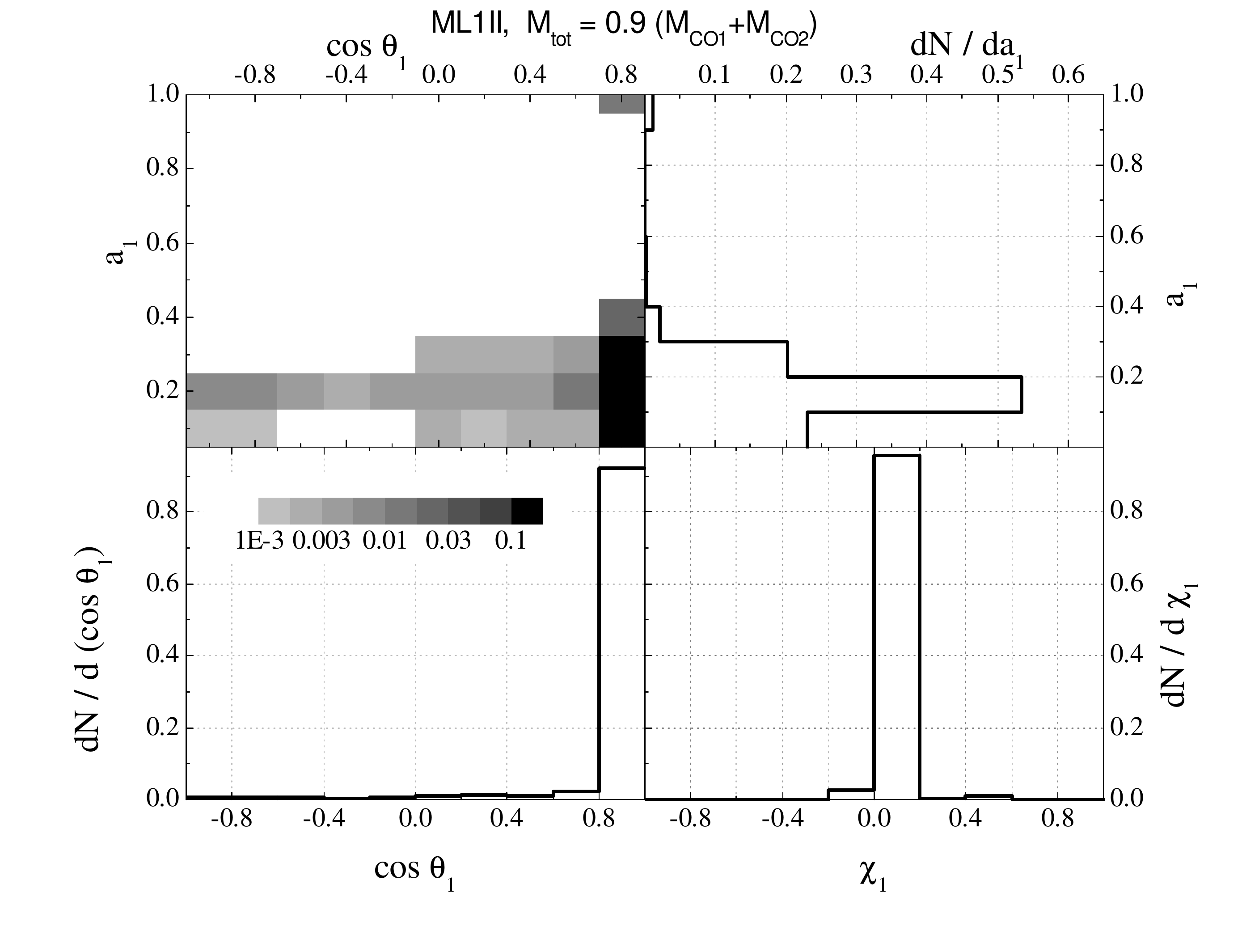}
\hfill
\includegraphics[width=0.49\textwidth]{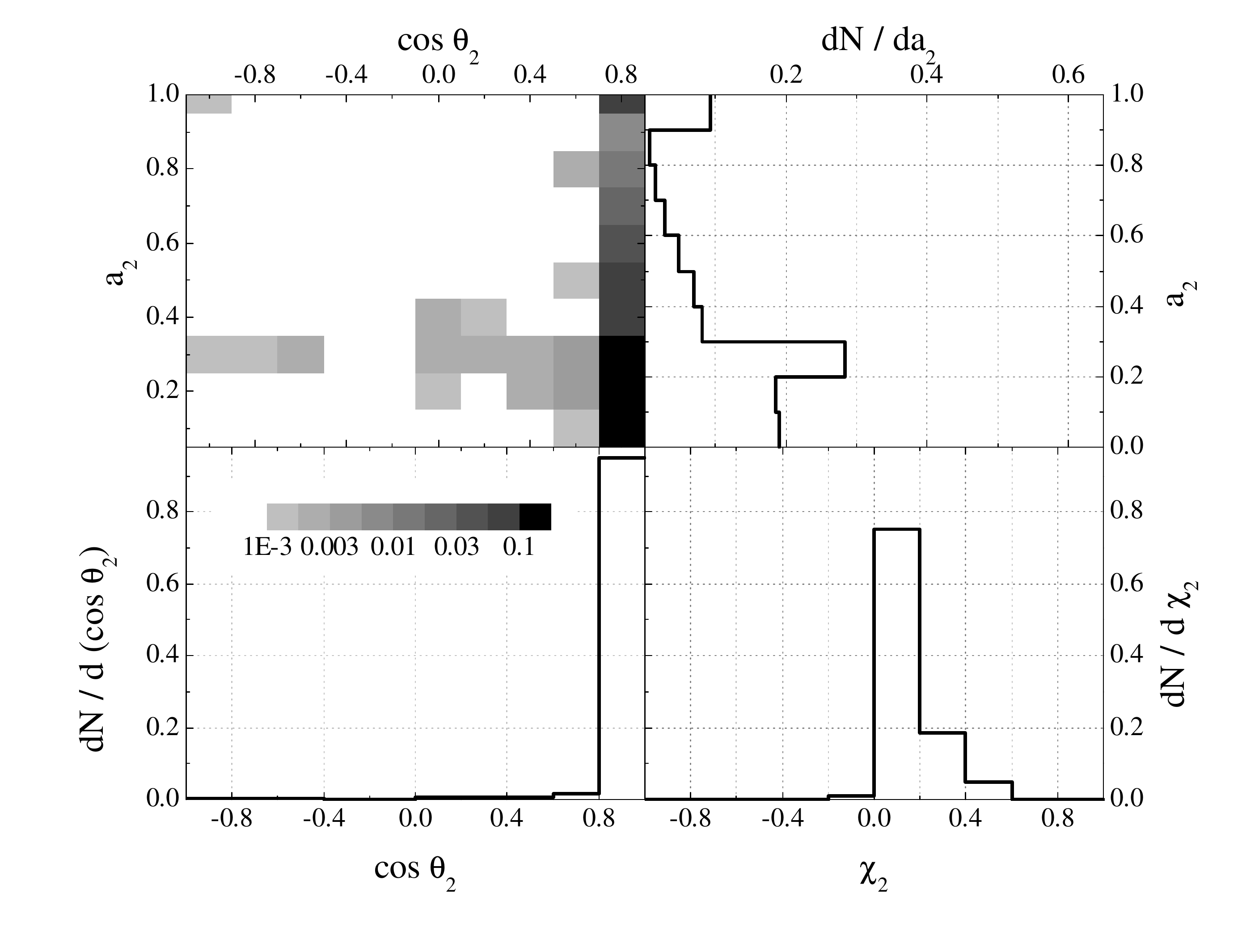}
\vfill
\includegraphics[width=0.49\textwidth]{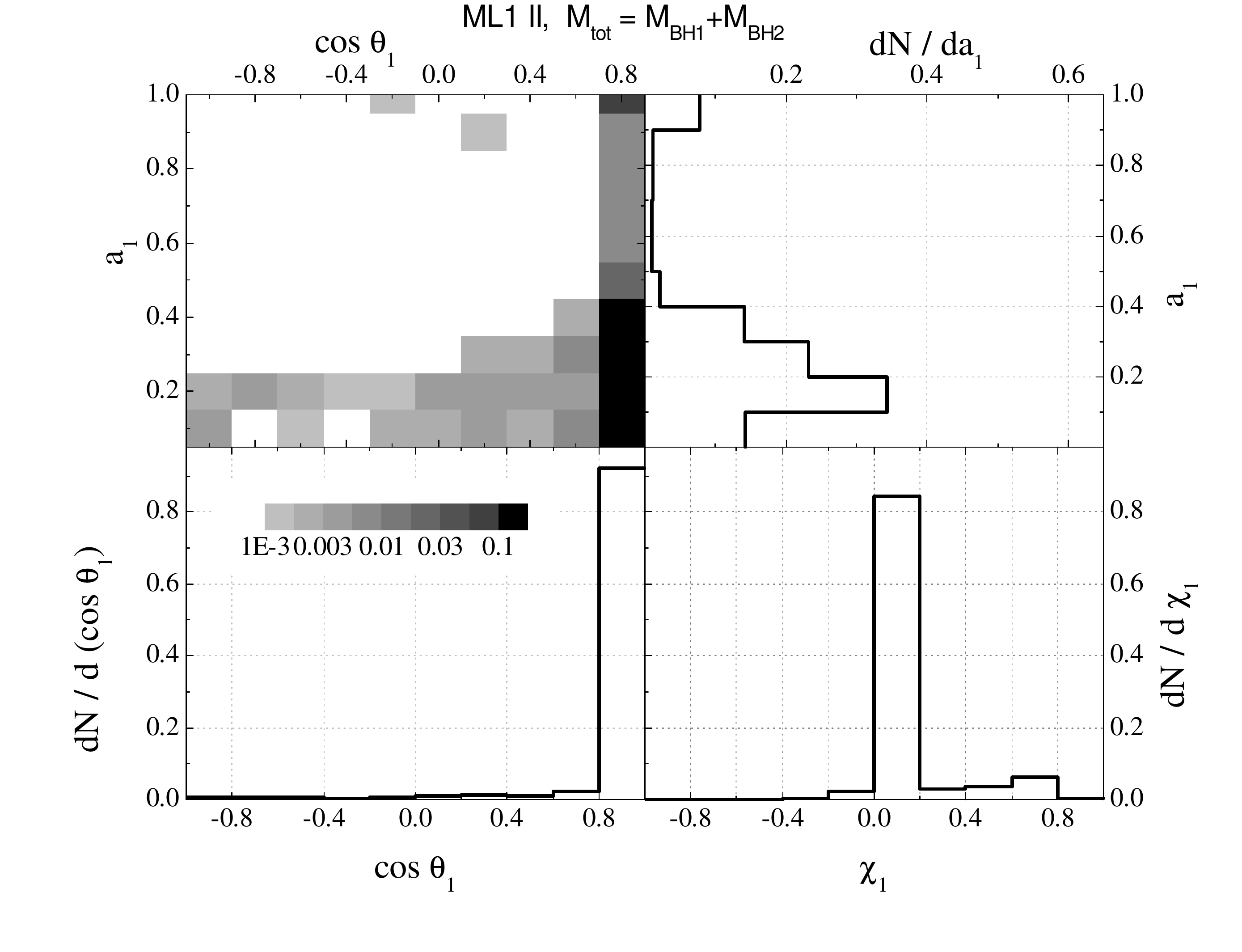}
\hfill
\includegraphics[width=0.49\textwidth]{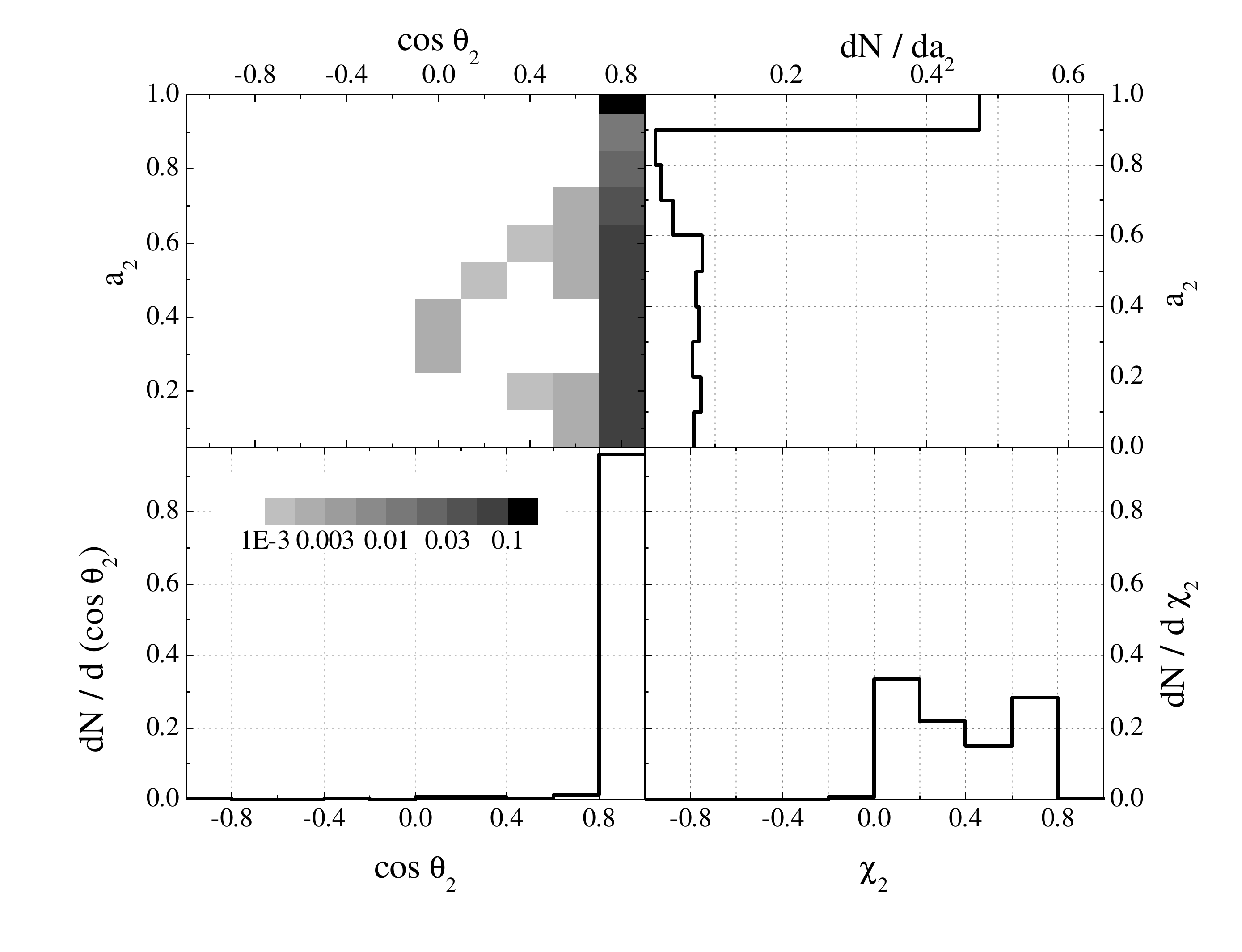}
\caption{Normalized distribution of spins $a_i$, misalignement angles $\cos\theta_i$ and mass-weighted spin projections on the orbital angular momentum $(M_i/M_\mathrm{tot})a_i\cos\theta_i$ of components of coalescing binary BHs. Shown is the case of ML1 stellar-wind mass loss, parallel initial binary component spins and   $\alpha_\mathrm{CE}=1$.  Panels in the upper and bottom rows correspond to BH formation model without and with envelope fallback, respectively (cf. 4th row-3d column in the upper and bottom panel of Fig. 3, respectively).  Left: BH from the primary component. Right: BH from the secondary component.}
\label{f:acost_par}
\end{center}
\end{figure*}

\begin{figure*}
\begin{center}
\includegraphics[width=0.49\textwidth]{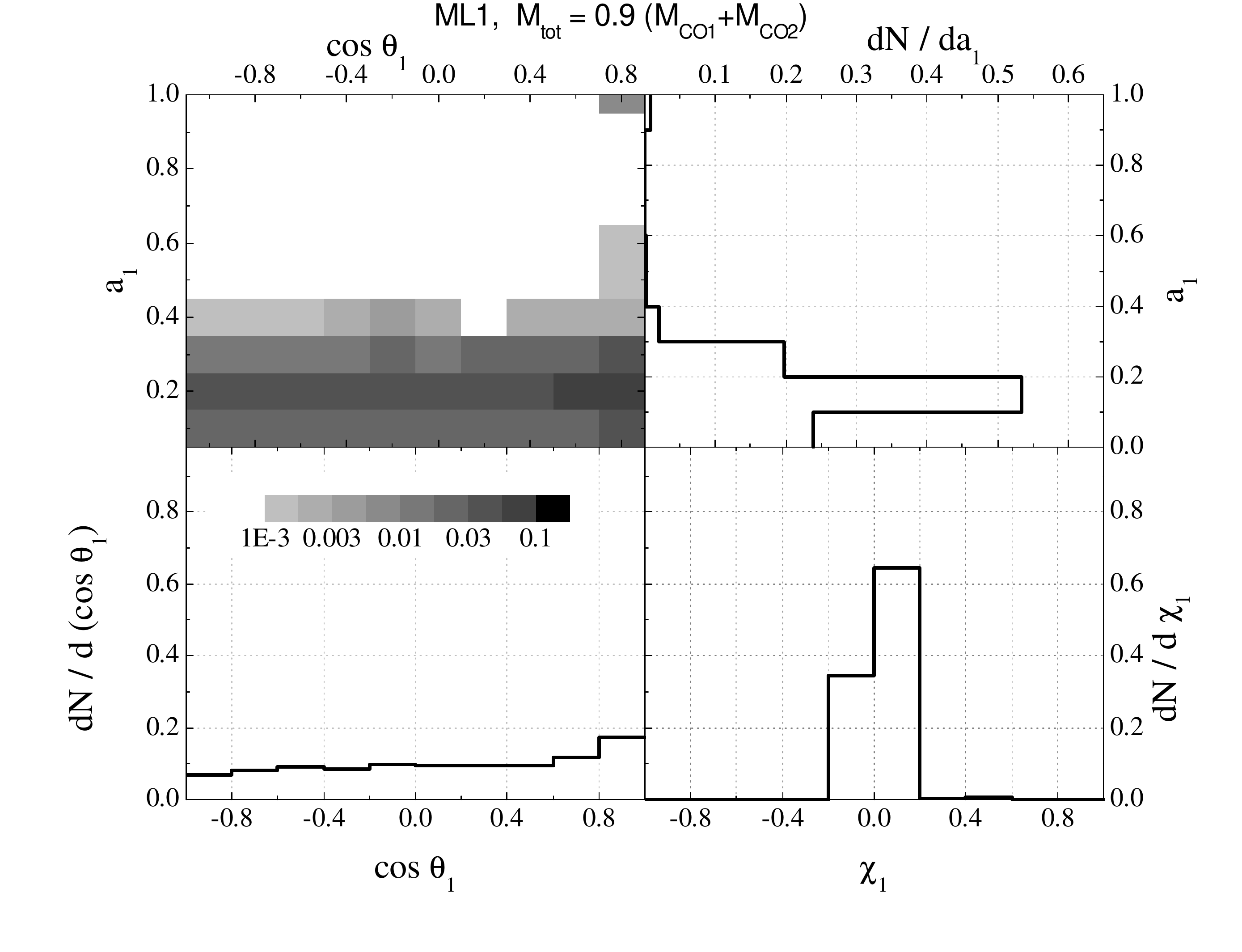}
\hfill
\includegraphics[width=0.49\textwidth]{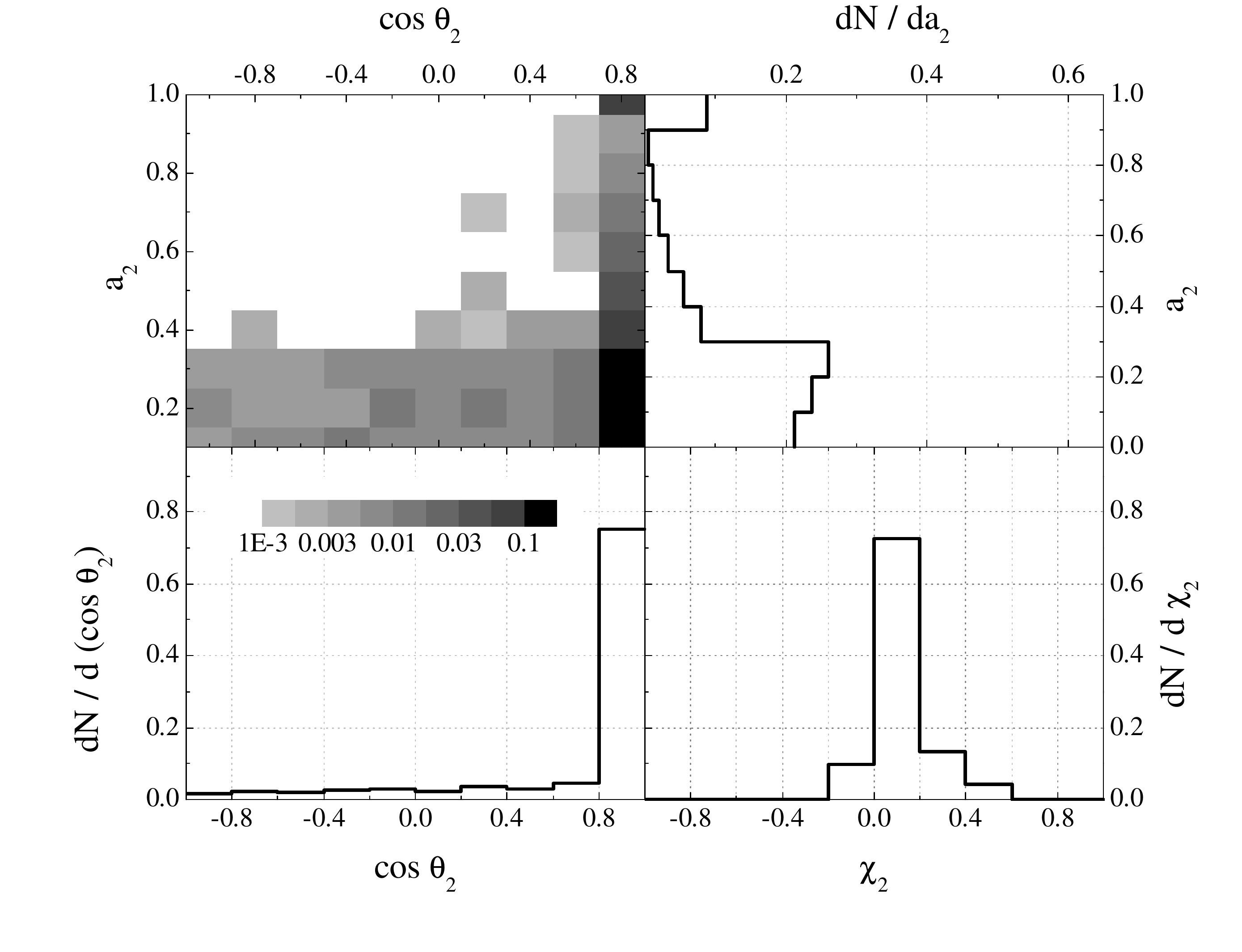}
\vfill
\includegraphics[width=0.49\textwidth]{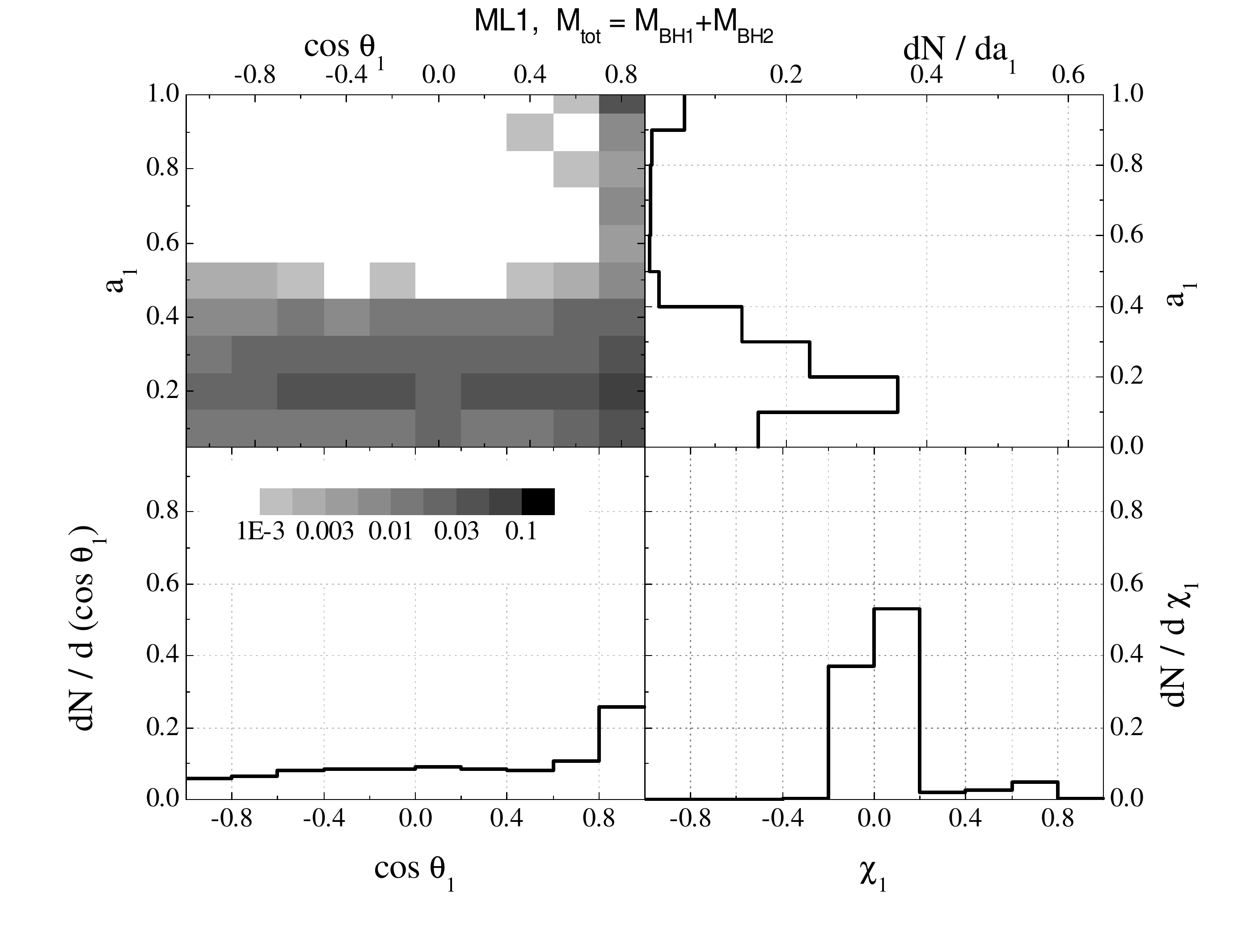}
\hfill
\includegraphics[width=0.49\textwidth]{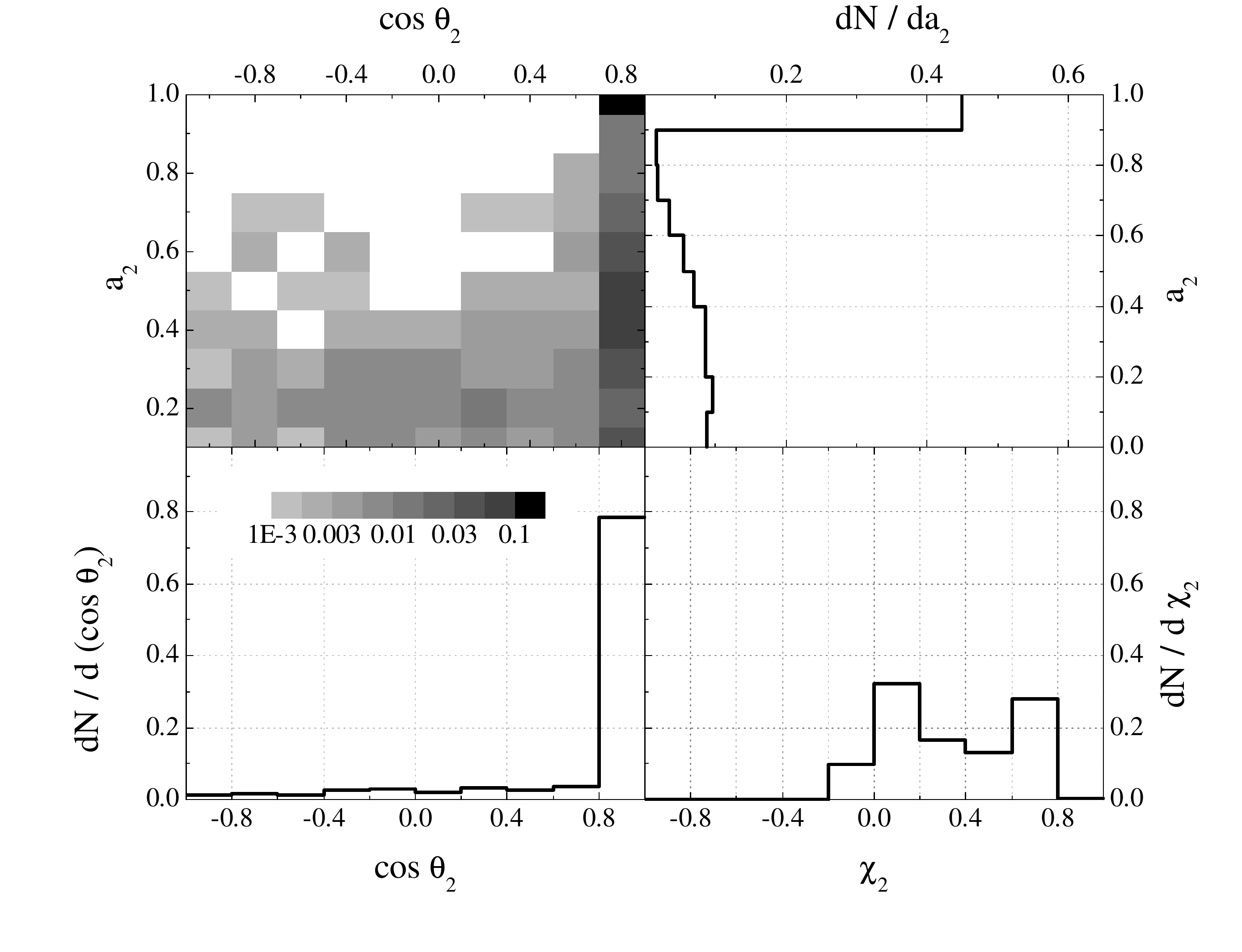}
\caption{The same as in Fig. \ref{f:acost_par} for randomly misaligned initial binary component spins (cf. 4th row-3d column in the upper and bottom panel of Fig. \ref{f:random_MLW1}, respectively).} 
\label{f:acost_ran}
\end{center}
\end{figure*}

\begin{figure*}
\begin{center}
\includegraphics[width=0.8\textwidth]{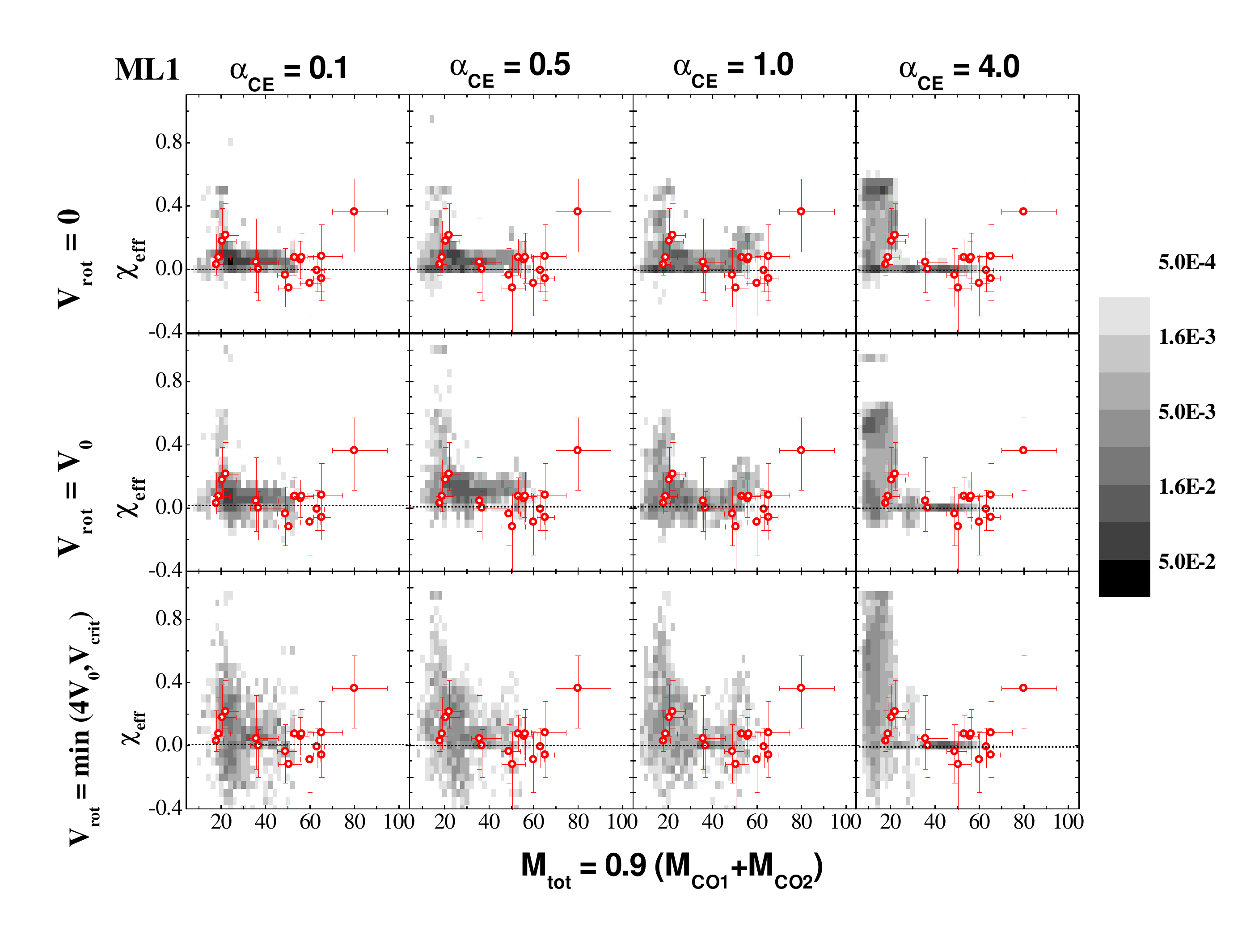}
\hfill
\includegraphics[width=0.8\textwidth]{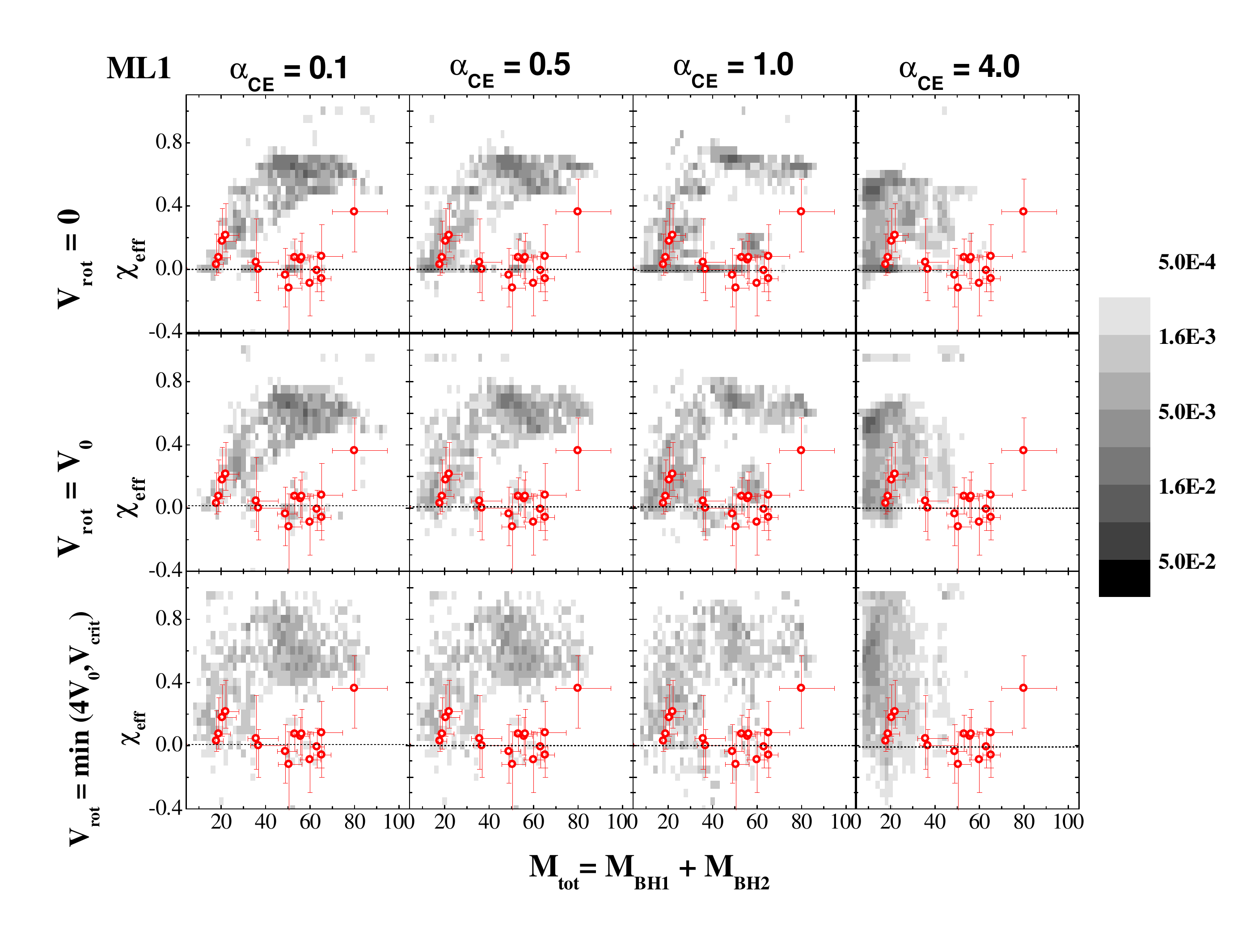}
\caption{$M_\mathrm{tot}-\chi_\mathrm{eff}$ distribution for different initial rotation velocities of the binary components. Results for the metallicity-dependent star-formation history \citep{2018arXiv180707659E}, $\alpha_\mathrm{CE}=0.5$, $\tau_c=5\times 10^5$~yrs.
Top:  the BH formation model without fallback from the outer envelope, $M_\mathrm{tot}=0.9(M_\mathrm{CO,1}+M_\mathrm{CO,2})$. Bottom: the BH formation model including the fallback from the outer envelope, $M_\mathrm{tot}=M_\mathrm{BH,1}+M_\mathrm{BH,2}$.} 
\label{f:vel}
\end{center}
\end{figure*}

\begin{figure*}
\begin{center}
\includegraphics[width=0.8\textwidth]{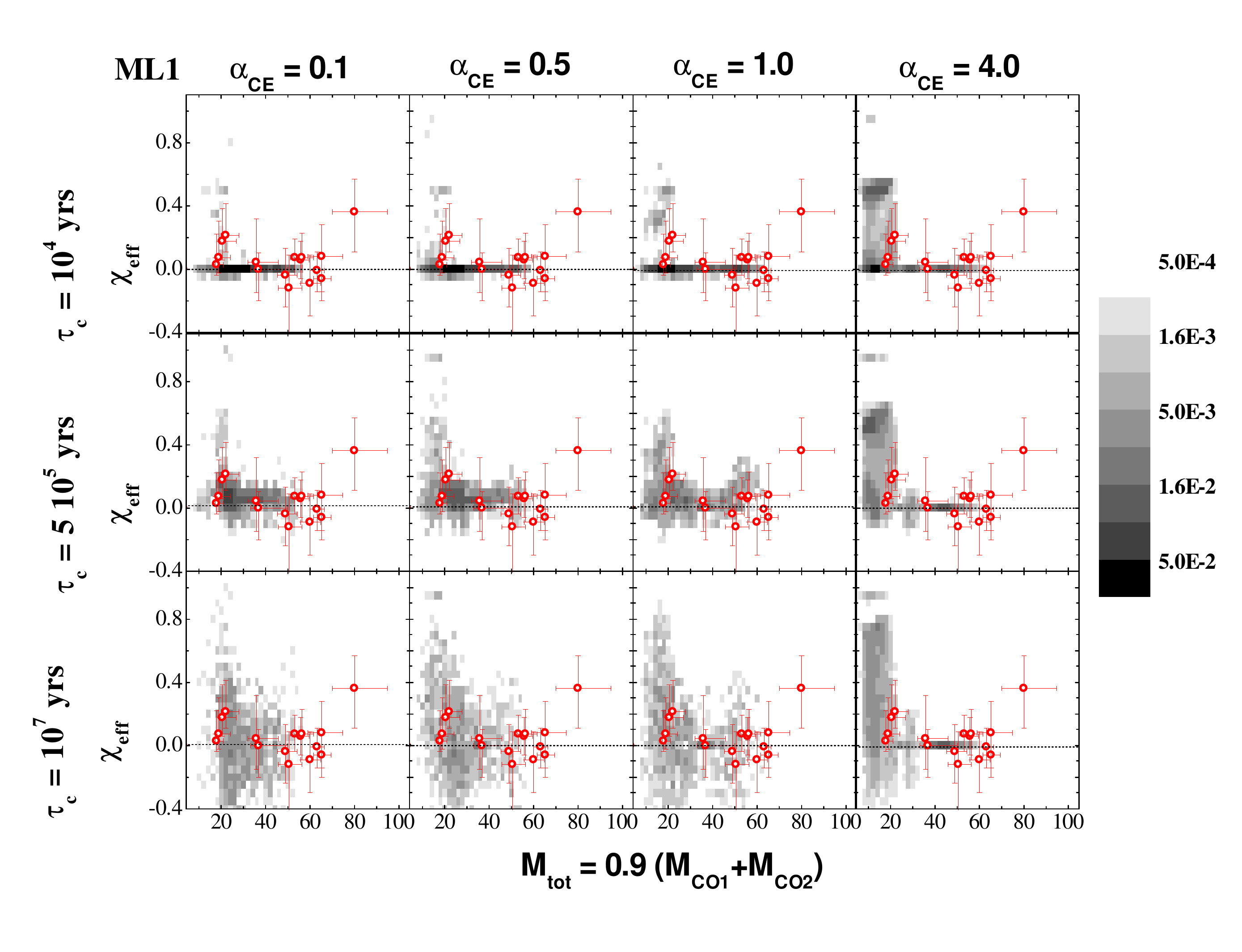}
\hfill
\includegraphics[width=0.8\textwidth]{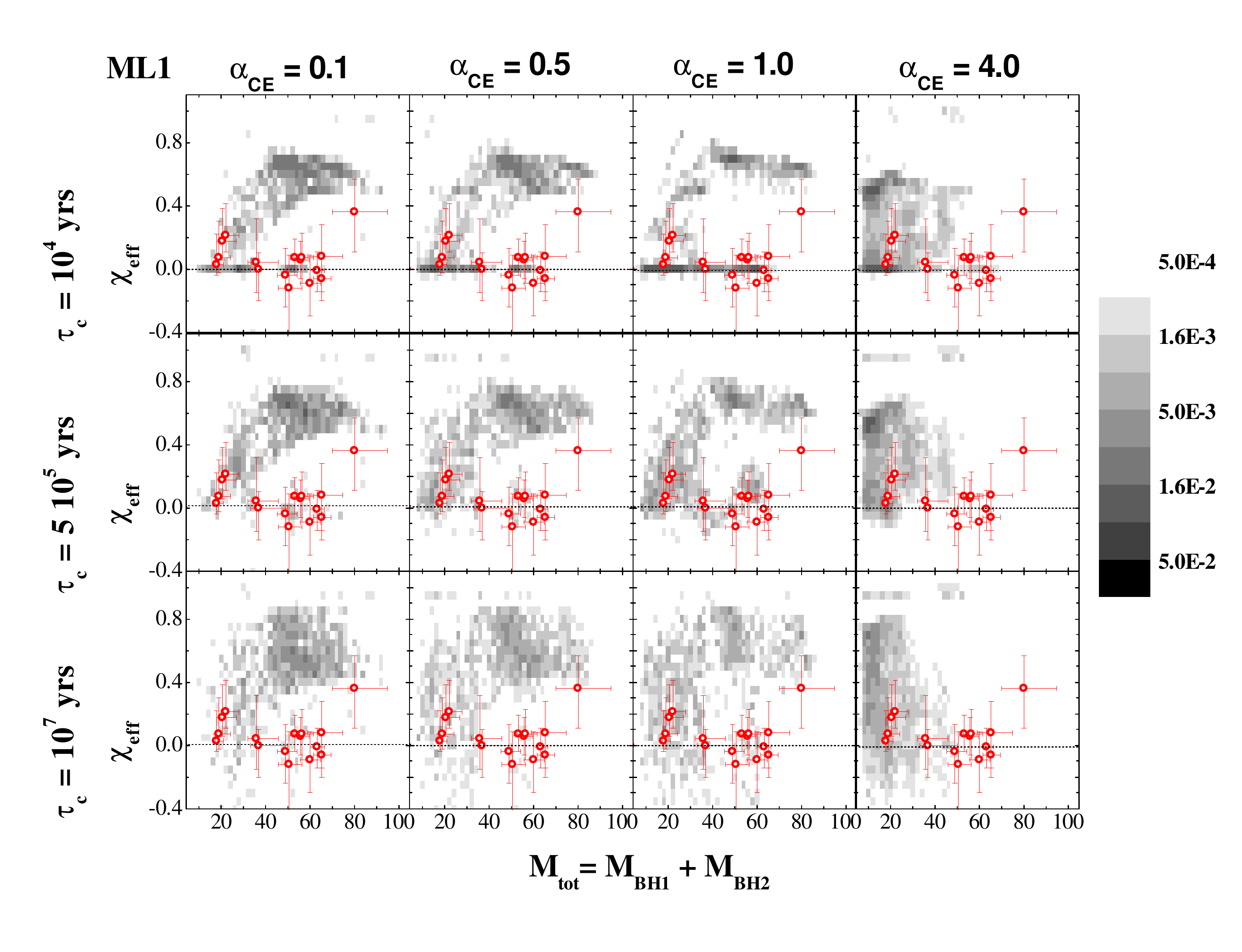}
\caption{$M_\mathrm{tot}-\chi_\mathrm{eff}$ distribution for different core-envelope rotational coupling time $\tau_c$. Results for the metallicity-dependent star-formation history \citep{2018arXiv180707659E}, $\alpha_\mathrm{CE}=0.5$, $v_0$ ( \Eq{e:vrot}).
Top:  the BH formation model without fallback from the outer envelope, $M_\mathrm{tot}=0.9(M_\mathrm{CO,1}+M_\mathrm{CO,2})$. Bottom: the BH formation model including the fallback from the outer envelope, $M_\mathrm{tot}=M_\mathrm{BH,1}+M_\mathrm{BH,2}$.
} 
\label{f:tau}
\end{center}
\end{figure*}

\subsection{BH spin misalignments}

The normalized distributions of BH spins $a$ and spin misalignments  in coalescing binary BHs, in terms of $\cos\theta$ relative to the orbital plane, for binaries convolved with the adopted metallicity-dependent star-formation history \citep{2018arXiv180707659E}, the common envelope efficiency $\alpha_\mathrm{CE}=1$, the ML1 stellar wind mass loss
are presented in Figs. \ref{f:acost_par} and 
\ref{f:acost_ran} for 
the initially coaligned and randomly misaligned spins of the binary components, respectively.
The BH formation model without and with the envelope fallback
are used to calculate the upper and bottom rows, respectively.

While in the both case the spin and mass distributions of the components ($dN/dM_\mathrm{BH1,2}$, Figs. \ref{f:JL0_MLW1}, \ref{f:random_MLW1}) look similar, for the initially aligned spins (Fig. \ref{f:acost_par}) the BH spins before the coalescence are aligned in most cases ($\cos\theta_{1,2}\simeq 1$). For randomly misaligned components the BH spins (Fig. \ref{f:acost_ran}) can be strongly misaligned , especially for the BH produced from the primary component (left bottom panels of the left column in Fig. \ref{f:acost_ran}). Indeed, for this component, the spin distribution almost 'remember' the initial random orientation of the primary spin because the spin of the primary stellar core of a rapidly evolving massive star has no time to tidally align with the orbital angular momentum before the first Roche lobe filling. In contrast, the spin of the secondary component (right column in Fig. \ref{f:acost_ran}) has time for the alignment, which gives rise to a pronounced peak in the BH spin orientation distribution at $\cos\theta=1$. Interestingly, in the initially misaligned spin case, there is a non-zero probability for the BH components to have counter-aligned spins. Note that such a possibility is still not excluded by observations of GW150914 due to GW  waveform degeneracy \citep{2018arXiv180302350C}.

Right bottom panels in Figs. \ref{f:acost_par} and 
\ref{f:acost_ran}
show the mass-weighted spin projections of BH components onto the orbital angular momentum, $(M_{1,2}/M_{tot})a_{1,2}\cos\theta_{1,2}$, before the coalescence determining the effective binary spin $\chi_\mathrm{eff}$. 
It is seen that in
both BH formation models (without or with fallback from the envelope) the second BH mostly contributes to $\chi_\mathrm{eff}$, especially in the case of the BH formation with fallback (bottom rows in these Figures).

Note that a natal BH kick alone is able to produce the BH spin-orbit misalignment. Indication to the possible non-zero natal BH kick was suggested by a careful statistical analysis of the BH spin misalignments in LIGO binary BHs carried out by \cite{2018PhRvD..97d3014W}.

\subsection{Effect of the initial stellar rotation and core-envelope coupling}

The influence of different assumptions about the initial rotation velocities of the binary components is shown in Fig. \ref{f:vel}. This Figure presents the calculated distribution $M_\mathrm{tot}-\chi_\mathrm{eff}$ for binaries convolved with the adopted metallicity-dependent star-formation rate \citep{2018arXiv180707659E}, different $\alpha_\mathrm{CE}$, the core-envelope rotational coupling time $\tau_c=5\times 10^5$~yrs and the ML1 stellar wind mass-loss model. The initial rotation of the binary components were calculated for equatorial velocities ranging from 0 to maximum possible rotation $v_\mathrm{crit}$ corresponding to the limiting break-up equatorial velocity of a rigidly rotating star. 

The effect of the different core-envelope time $\tau_c$ is shown for the same fiducial parameters but for the fixed initial velocity law \Eq{e:vrot} in Fig. \ref{f:tau}. Upper and bottom panels of both Figures correspond to the BH formation model without fallback and with fallback from the rotating envelope, respectively. 

Figs. \ref{f:vel} and \ref{f:tau} are almost identical (which is evident for the middle rows of each panels of the Figures that were calculated for the same parameters), but for plots with $v_0=0$
on the left panels, suggesting a degeneracy of the results with respect to the initial
rotational velocity $v_0$ of the components and core-envelope coupling efficiency 
parametrized by the time $\tau_c$. At first glance, this looks somewhat unexpected, but the analysis of individual evolutionary tracks in both cases suggests that it is a very effective tidal synchronization of the stellar envelopes at stages when the star fills its Roche lobe that determines the spin of the stellar C-O core (which we assume to collapse into BH on left panels). 

To see this, consider the upper panels of these figures (BH formation models without envelope fallback). The evolution of the angular momentum of the BH remnant is determined by two terms: the initial angular momentum of the C-O core and the change due to the core-envelope coupling. Increasing the initial star rotation from zero to maximum value (from top to bottom rows on these Figures) changes, correspondingly, the rotation of the C-O core for a given coupling time (Fig. \ref{f:vel}) thus widening the final $\chi_\mathrm{eff}$ distribution. On the other hand, at the given initial rotation of the star, the increase in the core-envelope coupling time (from an almost rigid coupling at low $\tau_c$ to independent rotation of the core and the envelope at large $\tau_c$, from top to bottom rows in Fig. \ref{f:tau})  decreases the angular momentum removal from the core, which also widens the resulting $\chi_\mathrm{eff}$ distribution. We remind that in our model the change in the core angular momentum is determined by the angular velocity difference between the stellar C-O core and the outer envelope.

In the case of BH formation with fallback (bottom panels of Figs. \ref{f:vel} and \ref{f:tau}), the final BH spin is mostly determined by the angular momentum transferred by accretion from the envelope, and the  evolution of the angular velocity of the envelope depends on the tidal synchronization at the Roche-lobe filling stages. The effect of a short tidal synchronization is clearly visible for BH spins calculated for less effective common envelopes with $\alpha_\mathrm{CE}=4$ (right columns): in this case, the most rapidly spinning BHs are formed from the tightest binaries prior to the beginning of the CE stage, which have the most rapidly rotating envelopes.

\section{Discussion and conclusions}

In the present paper, we have calculated the possible distribution of the total masses, 
$M_\mathrm{tot}$, and effective spins, $\chi_\mathrm{eff}$, of coalescing binary black holes formed through 
the standard astrophysical channel during evolution of massive binary stars of different metallicity. 
We have used a modified version of the open-access \textsc{bse} population synthesis code \citep{2000MNRAS.315..543H,2002MNRAS.329..897H}, 
to which we added the description of evolution of zero-metal Population III stars \citep{Kinugawa2014} 
and the treatment of the stellar core rotation in two-zone approximation as described in 
\citep{2016MNRAS.463.1642P}. To compare the results of calculations with the observed distributions of total masses $M_\mathrm{tot}$ and effective spins $\chi_\mathrm{eff}$, we have convloved the results of calculations for different metallicites with the metallicity-dependent star-formation rate history presented in \cite{2018arXiv180707659E} (see \Eq{e:sfr1} and \Eq{e:sfr2}). For completeness, we have calculated the evolution of zero-metallicity Population III stars \citep{Kinugawa2014} (Fig. \ref{f:Z0}).

Our calculations suggest that the effective spin $\chi_\mathrm{eff}$ of binary BH 
produced from massive binary star evolution (the standard astrophysical formation scenario for coalescing binary BHs) 
can be distributed in a wide range 
(Figs. \ref{f:random_MLW1}-\ref{f:random_MLW2}). 
The assumed BH formation model -- either without fallback from the outer rotating envelope of the collapsing star, when the total mass of the BH binary is determined by the mass of the stellar C-O core prior to the collapse, $M_\mathrm{tot}=0.9(M_\mathrm{CO,1}+M_\mathrm{CO,2})$, or with an account of the fallback from the outer envelope with $M_\mathrm{tot}=M_\mathrm{BH,1}+M_\mathrm{BH,2}$, where BH masses are calculated using the model of \citep{2012ApJ...749...91F} (Fig. \ref{f:mzams_mbh}) -- is found to mostly affect the final effective spin of the formed BH binary (cf. last rows in the upper and bottom panels of Figs.  \ref{f:JL0_MLW1}, \ref{f:random_MLW1}, \ref{f:random_MLW2}, \ref{f:Z0}. The increase in the BH spin during the BH formation with fallback has been independently confirmed by model calculations \citep{2018arXiv180501269S}. 
 
The second important assumption of our model calculations is the initial alignment of spins of the binary components. Initially misaligned binary components even with an account of tidal interaction during evolution give rise to misaligned BH
 spins and in some cases to negative effective spin parameter $\chi_\mathrm{eff}$ of the coalescing binary BHs. Some BH spin misalignment can also be 
 produced for initially coaligned spins due to possible natal BH kicks.

Other uncertainties, including the common envelope efficiency parameter $\alpha_\mathrm{CE}$ and the stellar wind mass-loss model for massive stars, initial rotational velocities of the binary components and effective core-envelope rotational coupling time, which we varied in 
the present calculations, have less strong effect on the results.

It is important to note that the spin of the secondary BH mostly contributes to the effective spin $\chi_\mathrm{eff}$ (see Figs. \ref{f:acost_par}, \ref{f:acost_ran}, the right column). This conclusion is independently confirmed by the recent calculations by the Geneva group \citep{2018arXiv180205738Q}. 

\begin{figure*}
\begin{center}
\includegraphics[width=0.49\textwidth]{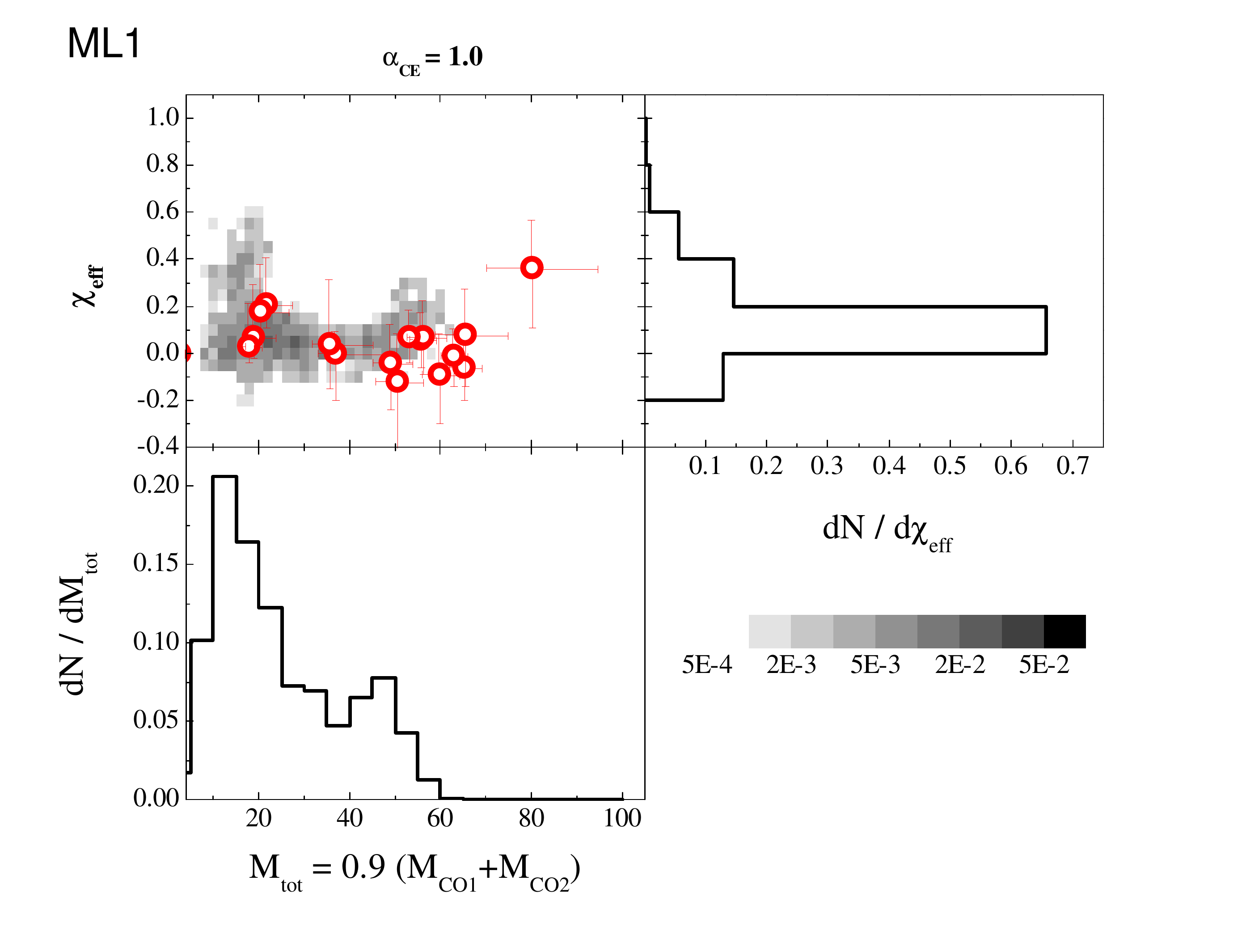}
\includegraphics[width=0.49\textwidth]{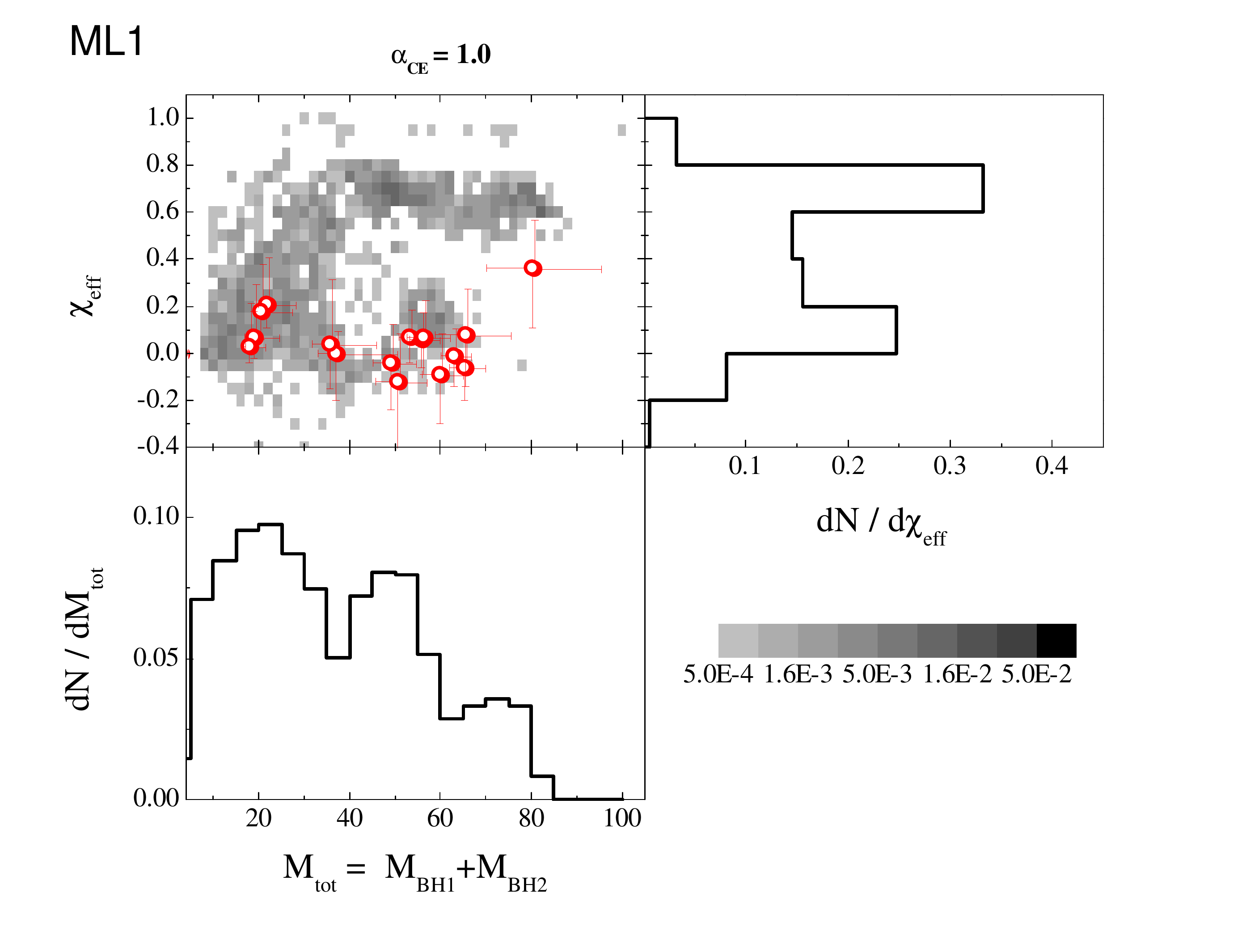}
\caption{
Normalized distribution of total mass -- effective spin ${M_\mathrm{tot}}-\chi_\mathrm{eff}$ of coalescing binary BH components. Shown is the case of ML1 stellar-wind mass loss,  randomly misaligned initial binary component spins and   $\alpha_\mathrm{CE}=1$. Left: BH formation model without fallback from the stellar envelope, $M_\mathrm{tot}=0.9(M_\mathrm{CO,1}+M_\mathrm{CO,2})$ (cf. 4th row-3d column in the upper panel of Fig. \ref{f:random_MLW1}). 
Right: BH formation model with fallback from the envelope, $M_\mathrm{tot}=M_\mathrm{BH,1}+M_\mathrm{BH,2}$ (cf. 4th row-3d column in the bottom panel of Fig. \ref{f:random_MLW1}).}
\label{f:2d_1d}
\end{center}
\end{figure*}

The inspection of the calculated model BH-BH binaries in Fig. \ref{f:random_MLW1}  suggests that the observed location of detected LIGO sources on the $M_\mathrm{tot}-\chi_\mathrm{eff}$ plane but the heaviest one, GW170729, can fall simultaneously within the calculated range of total masses and effective spins. To see this more clearly, we plot the seemingly most likely models (4th row -3d column in the upper and bottom panel of theis figure) as 1-d distributions separately in Fig. \ref{f:2d_1d}. It is seen that for the BH formation from the CO-core without addition from the surrounding envelope of the collapsing star, the total mass range does not cover the heaviest source, GW170729 (left panel). The allowance for additional fallback from the stellar envelope could reproduce the correct mass range but results in a much wider range of the effective spin $\chi_\mathrm{eff}$ of the coalescing binary BHs (right panel). Apparently, a more refined treatment of BH formation is required to reproduce simultaneously masses and spins of all observed so far LIGO BH-BH binaries, or a mixture of their formation channels should be involved.

Clearly, the calculation of the effective spins of coalescing binary BHs is subject to many uncertainties, which we tried to take into account in the present study. These include: (1) the initial stellar rotation, (2) the treatment of the angular momentum  transport in the star before the collapse, (3) the description of the mass loss, (4) the  calculation of BH spin during the collapse of a rotating star (see \Eq{e:a}). The assumption of the angular momentum conservation of the collapsing core appears to be safe, although the fraction of matter and angular momentum during the possible fallback is less reliable. However, connection of some long gamma-ray bursts with supernovae \citep{2012grb..book..169H} supports the collapsar model \citep{2006ARA&A..44..507W} involving rapidly rotating BHs from core collapses of massive stars and accompanied by mass ejection. Presently, efforts are being made to search for possible massive progenitors of type Ic supernovae (see, e.g., \cite{2018arXiv180301050V} and references therein), and failed supernovae \citep{2017MNRAS.468.4968A}, but the results are not fully conclusive.

In our calculations we have also taken into account a possible natal kick during the BH formation. This assumption remains 
model-dependent, but can be used to produce BH spin misalignments in the frame of the standard production channel of coalescing binary BHs from massive binary stars \citep[e.g.][and references therein]{2018PhRvD..97d3014W}. Note here that the BH kick law we used in our population synthesis calculations, \Eq{e:vkick}, is only one among many possible, which is difficult to specify at present. Moreover, allowing for off-center random kicks during the compact object formation could change the angular momentum as well \citep[see e.g.][for the case of the neutron star rotation]{1998Natur.393..139S,1998AstL...24..568P}. Clearly, for the BH case this issue remains open and requires further studies.

Our calculations suggest (see Fig. \ref{f:vel} and \ref{f:tau}) that there is a degeneracy between the initial rotation velocity of the binary components and the core-envelope coupling efficiency: the evolution of initially more rapidly rotating components and the evolution of mildly rotating (or even initially non-rotating) stars with less strong core-envelope coupling produce similar final
effective BH spin distributions. We also find that the fallback 
of matter from rotating envelope during BH formation 
always leads to higher effective BH spins, almost independently of the initial rotational velocity of the components $v_0$ and the core-envelope coupling time $\tau_\mathrm{c}$.

To facilitate the comparison with other recent population synthesis studies 
(e.g. \citealt{2018MNRAS.474.2959G}), we also computed the relative number of merging binary BHs per unit mass defined as $X=N_\mathrm{DBH}/\Sigma M_\mathrm{i}$, where $\Sigma M_\mathrm{i}$ is the total initial mass of stars calculated in each run with adopted distributions of masses and mass ratios (see Section 3). The results are listed in Table \ref{t:X} for different chemical compositions and parameters $\alpha_\mathrm{CE}$. Generally, our results agree 
with those calculations (cf. Table 3 in \cite{2018MNRAS.474.2959G}) because we have used very similar assumptions on the binary star evolution 
and BH formation but the description of the common envelope stage (those authors fixed both $\alpha_\mathrm{CE}$ and $\lambda$ parameters, while we 
have explicitly calculated the binding energy of the stellar envelope as described in \citealt{2011ApJ...743...49L}). 
The results are presented for two stellar wind mass loss models ML1 and ML2, which are identical for zero-metallicity stars (no wind mass loss was assumed) and are almost indistinguishable for solar metallicity stars.

\begin{table}
{\scriptsize
\centering
 \caption{The number of merging binary BHs per unit mass for different stellar metallicities, the common envelope efficiencies and stellar wind models \vspace{-1mm} \label{t:X}}
  \begin{tabular}{ll|ll}
 Z & $\alpha_\mathrm{CE} $ &   $X_{ML1}[M_\odot^{-1}]$ & $X_{ML2}[M_\odot^{-1}]$  \\
 \hline 
   0  & 0.1  &   2.2E-06 &   2.2E-06 \\
   0  & 0.5  &   2.2E-06 &   2.2E-06\\
   0  & 1.0  &   6.7E-06 &   6.7E-06\\
   0  & 4.0  &   4.8E-04 &   4.8E-04 \\
 \hline 
   0.0002  & 0.1  &   2.5E-06  &   8.6E-07\\
   0.0002  & 0.5  &   7.9E-06 &   4.1E-06\\
   0.0002  & 1.0  &   2.5E-05 &   2.6E-05\\
   0.0002  & 4.0  &   1.4E-04 &   1.1E-04\\
 \hline 
   0.002  & 0.1  &   2.4E-05 &   1.1E-05 \\
   0.002  & 0.5  &   2.8E-05 &   1.1E-05\\
   0.002  & 1.0  &   1.6E-05 &   5.9E-06\\
   0.002  & 4.0  &   6.3E-05 &   5.1E-05\\
 \hline 
   0.02  & 0.1  &      1.3E-07&   1.3E-07 \\
   0.02  & 0.5  &      1.5E-07&   1.5E-07 \\
   0.02  & 1.0  &      2.6E-07&   2.6E-07 \\
   0.02  & 4.0  &      1.5E-06&   1.5E-06 \\
 \end{tabular}
 }
 \end{table}

Of course, the predictive power of 
multi-parametric population synthesis calculations should not be overestimated. In addition to the distribution of the total mass
$M_\mathrm{tot}$ and effective spin $\chi_\mathrm{eff}$ we consider in the present paper (with the reservations discussed above in Section \ref{s:lowspins}), the occurrence rate of double BH mergings can also be used to constrain their evolutionary formation channels 
(see, e.g., \cite{2013ApJ...779...72D,2016Natur.534..512B,2016MNRAS.461.3877D,2017arXiv170607053B,2018arXiv180901152R}, among others). 

Presently, there are different viable pathways of producing massive binary BHs that merge in the Hubble time.
They can be formed from low-metallicity massive field stars, primordial Population III remnants, can be a result of dynamical evolution in dense 
stellar clusters or even primordial black holes. 
It is not excluded that all channels contribute to the observed 
binary BH population. For example, the discovery of very massive ($M >  50 M_\odot$) Schwarzschild BHs would be difficult to reconcile with the
standard massive binary evolution \citep{2016A&A...594A..97B}, but can be naturally explained by primordial black holes \citep{2016JCAP...11..036B}. 

With the current LIGO sensitivity, the detection horizon of binary BH with masses around 30 $M_\odot$ 
reaches $\sim 700$ Mpc (ignoring possible strong gravitational lensing, see \cite{2018arXiv180205273B}). So far the statistics of binary BH merging rate as a function of BH mass as inferred from 
reported LIGO events is consistent with a power-law dependence, $dR/dM\sim M^{-2.5}$ \citep{2017arXiv170203952H}, which does not contradict the general power-law behavior of the stellar mass function. 
Our calculations presented in this paper confirm that presently 
the formation of LIGO coalescing binary black holes can be explained in the frame of the standard astrophysical formation scenario, but the discovery of a very massive BH-BH binary with large and possible negative effective spin may require additional formation channels to these extreme objects.

Clearly, the increased statistics of BH masses and spins inferred from GW observations of 
binary BH mergings will be helpful to distinguish between the possible binary BH populations which can be formed at different stages of the evolution of the Universe.

\section*{Acknowledgements} 

The authors thank the anonymous referee for constructive criticism and useful notes. KP acknowledges the support from RSF grant 16-12-10519. AK acknowledges support from RSF grant 14-12-00146 (modification of the \textsc{bse} code) and the M.V. Lomonosov Moscow
State University Program of Development (Scientific School SAI MSU).

\bibliographystyle{mnras}
\bibliography{bh}

\appendix
\section{Summary of different types of evolutionary tracks}

Here we present in the table form a summary of different evolutionary tracks leading to the formation of coalescing binary BHs (Table A1). Columns: $Z$ -- stellar metallicity; $\alpha_\mathrm{CE}$ -- common envelope efficiency; CE1(2) -- stages of the components (primary -- star 1, secondary -- star 2) at the beginning of the first (second, if happens) CE stage (MS -- main sequence, HeMS -- helium main sequence; HeHG -- helim star Hertzspring gap;  CHeB -- core helium burning, RSG -- red super giant, BH -- black hole); $\frac{N}{N_\mathrm{DBH}}$ -- fraction of this type of tracks among all DBH binaries for this $Z$ and $\alpha_\mathrm{CE}$; $X[M_\odot^{-1}]$ -- the fraction of coalescing binary BH per total mass of calculated binaries;  $\langle M_\mathrm{tot}\rangle $ -- mean total mass of the coalescing BH binaries; $\langle \chi_\mathrm{eff}\rangle $, $\langle a_1\rangle $, $\langle a_2\rangle $ -- mean effective spin of the BH binary and individual BH spins at the merging, respectively, $\langle \cos\theta_{1,2}\rangle $ -- mean misalgnment angle cosines. $\langle M_\mathrm{tot}\rangle $, $\langle \chi_\mathrm{eff}\rangle $, $\langle a_{1,2}\rangle $, $\langle \cos\theta_{1,2}\rangle $ are shown for two BH formation models: with fallback from
the envelope when $M_\mathrm{tot}=M_\mathrm{BH1}+M_\mathrm{BH2}$, and without fallback, $M_\mathrm{tot}=0.9(M_\mathrm{CO1}+M_\mathrm{CO2})$.

The tracks were obtained in population synthesis runs of 1,000,000
binaries with given metallicity $Z$ and common envelope efficiency $\alpha_\mathrm{CE}$, with fixed initial parameter distributions (see Section \ref{s:BSE}) and other evolutionary parameters described in the main text: ML1 stellar wind mass loss model, the core-envelope coupling time $\tau_c=5\times 10^5$~yrs, the initial rotation of the components with velocity $v_0$ (\Eq{e:vrot}), and random initial spin misalgnment of the binary components.

\begin{table*}
 \centering
 {\scriptsize
 \caption{Merging double BHs in population synthesis run of 1,000,000 binaries with given $Z$ and $\alpha_\mathrm{CE}$ for two different BH formation models (see text) \vspace{-2mm} \label{tab:tab2}}
  \begin{tabular}{ll|ll|ll|ll|llllll|llllll}
 \hline 
 &  & \multicolumn{2}{c}{CE1} & \multicolumn{2} {c}{CE2} & & &  \multicolumn{6} {c}{$M_\mathrm{tot}=M_\mathrm{BH1}+M_\mathrm{BH2}$} & \multicolumn{6} {c}{$M_\mathrm{tot}=0.9(M_\mathrm{CO1}+M_\mathrm{CO2})$} \\
 Z & $\alpha_\mathrm{CE} $ &star 1 & star 2 & star 1 & star 2 & $\frac{N}{N_\mathrm{DBH}}$ & $X[M_\odot^{-1}]$ &  $\langle M_\mathrm{tot}\rangle $   & $\langle \chi_\mathrm{eff}\rangle $ & $\langle a_1\rangle $ & $\langle a_2\rangle $ & $\langle \cos{\theta_1}\rangle $ & $\langle \cos{\theta_2}\rangle $ &  $\langle M_\mathrm{tot}\rangle $   & $\langle \chi_\mathrm{eff}\rangle $ & $\langle a_{1}\rangle $ & $\langle a_{2}\rangle $ & $\langle \cos{\theta_1}\rangle $ & $\langle \cos{\theta_2}\rangle  $ \\ 
 \hline 
 \hline 
   0  & 0.1  & --    & --    & --    & --    &      1.00  &   2.2E-06  &      46.6  &      0.53  &      0.93  &      0.92  &      0.58  &      0.57  &      21.3  &      0.26  &      0.84  &      0.77  &      0.18  &      0.18 \\
 \hline 
   0  & 0.5  & --    & --    & --    & --    &      1.00  &   2.2E-06  &      46.6  &      0.47  &      0.90  &      0.91  &      0.53  &      0.53  &      21.4  &      0.26  &      0.85  &      0.77  &      0.20  &      0.20 \\
 \hline 
   0  & 1.0  & --    & --    & --    & --    &      0.33  &   2.2E-06  &      45.8  &      0.51  &      0.91  &      0.92  &      0.58  &      0.55  &      21.2  &      0.29  &      0.87  &      0.80  &      0.22  &      0.22 \\
   0  & 1.0  & RSG   & CHeB  & --    & --    &      0.01  &   1.0E-07  &      68.3  &      0.31  &      1.00  &      0.00  &      0.40  &      1.00  &      38.3  &      0.00  &      0.00  &      0.00  &      0.36  &      0.36 \\
   0  & 1.0  & BH    & RSG   & --    & --    &      0.65  &   4.4E-06  &      91.5  &      0.53  &      0.13  &      1.00  &      0.07  &      1.00  &      41.0  &      0.06  &      0.52  &      0.27  &      0.07  &      0.07 \\
 \hline 
   0  & 4.0  & CHeB  & MS    & --    & --    &      0.06  &   3.1E-05  &      40.8  &      0.38  &      0.52  &      0.86  &      0.29  &      0.95  &      38.2  &      0.10  &      0.00  &      0.32  &      0.17  &      0.17 \\
   0  & 4.0  & CHeB  & CHeB  & --    & --    &      0.08  &   4.1E-05  &      45.8  &      0.50  &      0.56  &      0.53  &      0.68  &      0.86  &      46.0  &      0.00  &      0.00  &      0.00  &      0.66  &      0.66 \\
   0  & 4.0  & RSG   & MS    & --    & --    &      0.42  &   0.0002  &      59.3  &      0.24  &      0.34  &      0.61  &      0.18  &      0.93  &      46.1  &      0.14  &      0.36  &      0.51  &     -0.04  &     -0.04 \\
   0  & 4.0  & RSG   & CHeB  & --    & --    &      0.33  &   1.6E-04  &      57.4  &      0.17  &      0.41  &      0.35  &      0.07  &      0.86  &      54.2  &     -0.01  &      0.36  &      0.00  &     -0.03  &     -0.03 \\
   0  & 4.0  & BH    & CHeB  & --    & --    &      0.01  &   5.5E-06  &      47.9  &      0.12  &      0.37  &      0.23  &      0.07  &      1.00  &      47.9  &      0.01  &      0.37  &      0.00  &      0.04  &      0.04 \\
   0  & 4.0  & BH    & RSG   & --    & --    &      0.06  &   3.0E-05  &      79.3  &      0.36  &      0.25  &      0.66  &      0.16  &      0.73  &      53.5  &      0.11  &      0.40  &      0.30  &      0.16  &      0.16 \\
 \hline 
 \hline 
   0.0002  & 0.1  & --    & --    & --    & --    &      0.14  &   3.4E-07  &      18.4  &      0.22  &      0.25  &      0.39  &      0.67  &      0.62  &      16.4  &      0.15  &      0.20  &      0.34  &      0.37  &      0.37 \\
   0.0002  & 0.1  & RSG   & RSG   & --    & --    &      0.07  &   1.7E-07  &      44.2  &     -0.01  &      0.15  &      0.22  &     -0.16  &      0.05  &      44.2  &     -0.01  &      0.15  &      0.22  &     -0.16  &     -0.16 \\
   0.0002  & 0.1  & BH    & RSG   & --    & --    &      0.79  &   2.0E-06  &      52.5  &      0.30  &      0.26  &      0.54  &      0.07  &      0.49  &      41.6  &      0.12  &      0.21  &      0.31  &      0.05  &      0.05 \\
 \hline 
   0.0002  & 0.5  & --    & --    & --    & --    &      0.05  &   3.8E-07  &      16.5  &      0.13  &      0.21  &      0.38  &      0.35  &      0.62  &      14.9  &      0.14  &      0.19  &      0.38  &      0.28  &      0.28 \\
   0.0002  & 0.5  & RSG   & RSG   & --    & --    &      0.12  &   9.4E-07  &      46.2  &      0.03  &      0.15  &      0.23  &      0.10  &      0.16  &      46.2  &      0.03  &      0.15  &      0.23  &      0.10  &      0.10 \\
   0.0002  & 0.5  & BH    & RSG   & --    & --    &      0.82  &   6.4E-06  &      51.9  &      0.31  &      0.26  &      0.60  &      0.10  &      0.65  &      45.3  &      0.20  &      0.22  &      0.46  &      0.08  &      0.08 \\
 \hline 
   0.0002  & 1.0  & --    & --    & --    & --    &      0.01  &   3.0E-07  &      15.8  &      0.19  &      0.22  &      0.41  &      0.51  &      0.73  &      14.6  &      0.18  &      0.20  &      0.41  &      0.33  &      0.33 \\
   0.0002  & 1.0  & RSG   & MS    & BH    & RSG   &      0.02  &   3.9E-07  &      44.8  &      0.41  &      0.19  &      0.83  &      0.32  &      0.84  &      32.8  &      0.04  &      0.19  &      0.14  &      0.29  &      0.29 \\
   0.0002  & 1.0  & RSG   & RSG   & --    & --    &      0.04  &   1.1E-06  &      47.2  &      0.01  &      0.17  &      0.25  &      0.01  &      0.13  &      47.2  &      0.01  &      0.17  &      0.25  &      0.01  &      0.01 \\
   0.0002  & 1.0  & BH    & CHeB  & --    & --    &      0.03  &   7.7E-07  &      26.8  &      0.29  &      0.64  &      0.47  &      0.29  &      0.84  &      17.2  &      0.19  &      0.22  &      0.47  &      0.25  &      0.25 \\
   0.0002  & 1.0  & BH    & RSG   & --    & --    &      0.89  &   2.3E-05  &      47.1  &      0.46  &      0.22  &      0.81  &      0.05  &      0.89  &      31.8  &      0.16  &      0.22  &      0.37  &      0.01  &      0.01 \\
 \hline 
   0.0002  & 4.0  & CHeB  & CHeB  & --    & --    &      0.04  &   6.0E-06  &      27.7  &      0.57  &      0.41  &      0.79  &      0.74  &      0.96  &      26.2  &      0.27  &      0.13  &      0.45  &      0.66  &      0.66 \\
   0.0002  & 4.0  & RSG   & CHeB  & --    & --    &      0.17  &   2.5E-05  &      35.2  &      0.52  &      0.54  &      0.83  &      0.51  &      0.89  &      25.7  &      0.24  &      0.23  &      0.66  &      0.00  &      0.00 \\
   0.0002  & 4.0  & BH    & CHeB  & --    & --    &      0.53  &   7.5E-05  &      22.6  &      0.45  &      0.22  &      0.84  &      0.35  &      0.95  &      20.6  &      0.33  &      0.13  &      0.63  &      0.32  &      0.32 \\
   0.0002  & 4.0  & BH    & CHeB  & BH    & HeHG  &      0.03  &   4.5E-06  &      10.5  &      0.52  &      0.12  &      0.99  &      0.75  &      0.99  &      10.3  &      0.51  &      0.12  &      0.99  &      0.70  &      0.70 \\
   0.0002  & 4.0  & BH    & RSG   & --    & --    &      0.20  &   2.8E-05  &      50.0  &      0.53  &      0.22  &      0.88  &      0.06  &      0.97  &      32.2  &      0.18  &      0.22  &      0.38  &      0.02  &      0.02 \\
 \hline 
 \hline 
   0.002  & 0.1  & CHeB  & CHeB  & --    & --    &      0.02  &   5.2E-07  &      34.2  &      0.28  &      0.35  &      0.55  &      0.29  &      0.53  &      29.6  &      0.19  &      0.14  &      0.44  &      0.29  &      0.29 \\
   0.002  & 0.1  & BH    & CHeB  & --    & --    &      0.07  &   1.6E-06  &      28.7  &      0.36  &      0.30  &      0.72  &      0.08  &      0.98  &      21.4  &      0.33  &      0.22  &      0.65  &      0.09  &      0.09 \\
   0.002  & 0.1  & BH    & RSG   & --    & --    &      0.90  &   2.1E-05  &      54.3  &      0.55  &      0.17  &      0.89  &      0.07  &      0.94  &      32.0  &      0.08  &      0.13  &      0.18  &      0.05  &      0.05 \\
 \hline 
   0.002  & 0.5  & CHeB  & CHeB  & --    & --    &      0.16  &   4.4E-06  &      19.9  &      0.30  &      0.33  &      0.43  &      0.62  &      0.71  &      17.8  &      0.27  &      0.26  &      0.38  &      0.56  &      0.56 \\
   0.002  & 0.5  & RSG   & CHeB  & --    & --    &      0.07  &   2.0E-06  &      29.4  &      0.40  &      0.51  &      0.23  &      0.75  &      0.53  &      19.9  &      0.06  &      0.13  &      0.18  &      0.06  &      0.06 \\
   0.002  & 0.5  & HeMS  & CHeB  & --    & --    &      0.01  &   3.6E-07  &      16.1  &      0.63  &      0.90  &      0.40  &      0.98  &      0.61  &      15.2  &      0.64  &      0.93  &      0.40  &      0.99  &      0.99 \\
   0.002  & 0.5  & BH    & CHeB  & --    & --    &      0.15  &   4.0E-06  &      24.5  &      0.32  &      0.25  &      0.61  &      0.26  &      0.89  &      21.0  &      0.19  &      0.20  &      0.40  &      0.22  &      0.22 \\
   0.002  & 0.5  & BH    & RSG   & --    & --    &      0.59  &   1.6E-05  &      55.3  &      0.57  &      0.17  &      0.90  &      0.07  &      0.95  &      33.2  &      0.09  &      0.13  &      0.19  &      0.05  &      0.05 \\
 \hline 
   0.002  & 1.0  & CHeB  & MS    & --    & --    &      0.02  &   2.5E-07  &      16.4  &      0.19  &      0.33  &      0.34  &      0.52  &      0.54  &      14.8  &      0.10  &      0.29  &      0.34  &      0.10  &      0.10 \\
   0.002  & 1.0  & CHeB  & CHeB  & --    & --    &      0.17  &   2.8E-06  &      22.9  &      0.14  &      0.26  &      0.36  &      0.38  &      0.29  &      20.8  &      0.06  &      0.14  &      0.21  &      0.29  &      0.29 \\
   0.002  & 1.0  & RSG   & MS    & BH    & CHeB  &      0.01  &   1.7E-07  &      22.9  &      0.69  &      0.66  &      1.00  &      0.76  &      0.99  &      15.0  &      0.32  &      0.13  &      1.00  &      0.18  &      0.18 \\
   0.002  & 1.0  & RSG   & CHeB  & --    & --    &      0.10  &   1.6E-06  &      33.9  &      0.44  &      0.60  &      0.18  &      0.77  &      0.37  &      22.3  &      0.03  &      0.14  &      0.12  &      0.06  &      0.06 \\
   0.002  & 1.0  & RSG   & RSG   & --    & --    &      0.01  &   1.7E-07  &      66.6  &      0.37  &      0.21  &      0.57  &     -0.19  &      0.66  &      44.6  &     -0.01  &      0.15  &      0.12  &     -0.26  &     -0.26 \\
   0.002  & 1.0  & HeMS  & CHeB  & --    & --    &      0.02  &   2.7E-07  &      14.7  &      0.64  &      0.84  &      0.52  &      0.96  &      0.75  &      14.1  &      0.66  &      0.86  &      0.52  &      0.99  &      0.99 \\
   0.002  & 1.0  & BH    & CHeB  & --    & --    &      0.28  &   4.5E-06  &      27.9  &      0.29  &      0.26  &      0.61  &      0.21  &      0.88  &      24.8  &      0.14  &      0.17  &      0.32  &      0.19  &      0.19 \\
   0.002  & 1.0  & BH    & CHeB  & BH    & HeHG  &      0.01  &   1.7E-07  &      25.9  &      0.57  &      0.31  &      1.00  &      0.31  &      1.00  &      20.1  &      0.55  &      0.26  &      1.00  &      0.31  &      0.31 \\
   0.002  & 1.0  & BH    & RSG   & --    & --    &      0.39  &   6.3E-06  &      59.5  &      0.51  &      0.20  &      0.76  &      0.08  &      0.95  &      41.5  &      0.12  &      0.12  &      0.24  &      0.07  &      0.07 \\
 \hline 
   0.002  & 4.0  & CHeB  & MS    & BH    & CHeB  &      0.03  &   1.8E-06  &      12.3  &      0.36  &      0.25  &      0.68  &      0.47  &      0.85  &      12.1  &      0.35  &      0.26  &      0.68  &      0.36  &      0.36 \\
   0.002  & 4.0  & CHeB  & CHeB  & --    & --    &      0.11  &   6.8E-06  &      29.4  &      0.39  &      0.47  &      0.56  &      0.57  &      0.56  &      28.0  &      0.12  &      0.17  &      0.21  &      0.54  &      0.54 \\
   0.002  & 4.0  & RSG   & MS    & BH    & CHeB  &      0.01  &   9.2E-07  &      19.0  &      0.30  &      0.41  &      0.20  &      0.75  &      0.52  &      14.0  &      0.07  &      0.12  &      0.20  &      0.30  &      0.30 \\
   0.002  & 4.0  & RSG   & CHeB  & --    & --    &      0.05  &   2.9E-06  &      26.1  &      0.25  &      0.31  &      0.36  &      0.60  &      0.56  &      22.3  &      0.08  &      0.12  &      0.26  &      0.13  &      0.13 \\
   0.002  & 4.0  & BH    & CHeB  & --    & --    &      0.76  &   4.8E-05  &      25.7  &      0.40  &      0.29  &      0.75  &      0.27  &      0.94  &      23.6  &      0.28  &      0.18  &      0.52  &      0.23  &      0.23 \\
   0.002  & 4.0  & BH    & RSG   & --    & --    &      0.03  &   1.6E-06  &      65.0  &      0.34  &      0.19  &      0.56  &      0.12  &      0.93  &      53.0  &      0.14  &      0.11  &      0.28  &      0.12  &      0.12 \\
 \hline 
 \hline 
   0.02  & 0.1  & RSG   & CHeB  & --    & --    &      0.02  &   2.7E-09  &      14.5  &      0.04  &      0.14  &      0.05  &      0.06  &      0.81  &       9.3  &      0.01  &      0.11  &      0.05  &     -0.24  &     -0.24 \\
   0.02  & 0.1  & BH    & CHeB  & --    & --    &      0.03  &   4.0E-09  &      19.3  &      0.17  &      0.10  &      0.38  &      0.09  &      0.90  &      11.4  &      0.20  &      0.10  &      0.39  &      0.27  &      0.27 \\
   0.02  & 0.1  & BH    & RSG   & --    & --    &      0.94  &   1.2E-07  &      23.4  &      0.10  &      0.17  &      0.17  &      0.45  &      0.46  &      12.8  &      0.08  &      0.17  &      0.14  &      0.34  &      0.34 \\
 \hline 
   0.02  & 0.5  & CHeB  & CHeB  & --    & --    &      0.09  &   1.4E-08  &      22.5  &      0.26  &      0.21  &      0.38  &      0.50  &      0.69  &      12.7  &      0.22  &      0.14  &      0.36  &      0.44  &      0.44 \\
   0.02  & 0.5  & RSG   & CHeB  & --    & --    &      0.19  &   2.9E-08  &      13.8  &      0.13  &      0.18  &      0.09  &      0.75  &      0.50  &       8.3  &      0.04  &      0.07  &      0.09  &      0.20  &      0.20 \\
   0.02  & 0.5  & RSG   & RSG   & --    & --    &      0.01  &   1.7E-09  &      15.7  &      0.04  &      0.09  &      0.10  &     -0.04  &      0.51  &      10.1  &      0.04  &      0.09  &      0.10  &     -0.07  &     -0.07 \\
   0.02  & 0.5  & BH    & CHeB  & --    & --    &      0.09  &   1.3E-08  &      18.7  &      0.21  &      0.10  &      0.48  &      0.55  &      0.64  &      11.2  &      0.21  &      0.11  &      0.48  &      0.49  &      0.49 \\
   0.02  & 0.5  & BH    & RSG   & --    & --    &      0.62  &   9.3E-08  &      23.6  &      0.10  &      0.17  &      0.19  &      0.38  &      0.38  &      13.0  &      0.08  &      0.17  &      0.17  &      0.32  &      0.32 \\
 \hline 
   0.02  & 1.0  & CHeB  & CHeB  & --    & --    &      0.20  &   5.1E-08  &      20.1  &      0.20  &      0.18  &      0.33  &      0.45  &      0.49  &      11.7  &      0.19  &      0.15  &      0.31  &      0.39  &      0.39 \\
   0.02  & 1.0  & RSG   & CHeB  & --    & --    &      0.21  &   5.5E-08  &      13.2  &      0.19  &      0.16  &      0.33  &      0.53  &      0.59  &       8.0  &      0.11  &      0.07  &      0.33  &      0.08  &      0.08 \\
   0.02  & 1.0  & BH    & CHeB  & --    & --    &      0.25  &   6.4E-08  &      18.9  &      0.30  &      0.12  &      0.80  &      0.46  &      0.87  &      11.2  &      0.31  &      0.11  &      0.80  &      0.43  &      0.43 \\
   0.02  & 1.0  & BH    & CHeB  & BH    & HeHG  &      0.02  &   6.4E-09  &      16.7  &      0.45  &      0.11  &      1.00  &     -0.06  &      1.00  &      10.4  &      0.44  &      0.11  &      1.00  &     -0.10  &     -0.10 \\
   0.02  & 1.0  & BH    & RSG   & --    & --    &      0.30  &   7.7E-08  &      23.7  &      0.11  &      0.12  &      0.23  &      0.22  &      0.56  &      12.9  &      0.06  &      0.12  &      0.14  &      0.19  &      0.19 \\
 \hline 
   0.02  & 4.0  & CHeB  & CHeB  & --    & --    &      0.25  &   3.7E-07  &      24.9  &      0.39  &      0.42  &      0.53  &      0.60  &      0.55  &      15.0  &      0.24  &      0.25  &      0.39  &      0.57  &      0.57 \\
   0.02  & 4.0  & BH    & CHeB  & --    & --    &      0.72  &   1.1E-06  &      27.7  &      0.04  &      0.19  &      0.12  &      0.13  &      0.37  &      14.9  &      0.03  &      0.16  &      0.09  &      0.12  &      0.12 \\
   0.02  & 4.0  & BH    & RSG   & --    & --    &      0.01  &   1.9E-08  &      22.6  &      0.12  &      0.12  &      0.21  &      0.40  &      0.79  &      12.6  &      0.07  &      0.09  &      0.15  &      0.27  &      0.27 \\
 \hline 
 \end{tabular}
 }
 \end{table*}

\bsp	
\label{lastpage}
\end{document}